\documentclass{aa}  
\usepackage[T1]{fontenc}
\usepackage[utf8]{inputenc}
\usepackage{amsmath, amssymb, mathtools,bm}		
\usepackage{siunitx}
\usepackage[english]{varioref}				
\usepackage{xcolor}
\usepackage{graphicx}
\usepackage{txfonts}
\usepackage{threeparttable}
\usepackage{booktabs}
\usepackage{diagbox}
\usepackage{multirow}	
\usepackage{acronym}
\usepackage[labelfont=bf, font=small]{caption} 	
\usepackage{comment}
\usepackage{xspace} 
\usepackage{float}  

\usepackage{natbib}
\usepackage[breaklinks=true]{hyperref}\hypersetup{hidelinks,colorlinks=true,linkcolor=blue,citecolor=blue, urlcolor=blue}

\newcommand{\msun}{\ensuremath{\rm ~M_{\odot}}\xspace}
\newcommand{\rsun}{\ensuremath{~R_{\odot}}\xspace}

\newcommand{\mdot}{\ensuremath{~\dot{M}}}
\newcommand{\mzams}{\ensuremath{M_\textup{ZAMS}}}

\newcommand{\ffrom}{\ensuremath{f_\textup{from}}\xspace}
\newcommand{\ffate}{\ensuremath{ f_\textup{fate}}\xspace}
\newcommand{\mproto}{\ensuremath{M_{\rm proto}}}
\newcommand{\mej}{\ensuremath{M_\textup{ej}}}
\newcommand{\mhe}{\ensuremath{M_\textup{He}}}
\newcommand{\mhecore}{\ensuremath{M_\textup{He-core}}}
\newcommand{\mstar}{\ensuremath{M_\textup{star}}}
\newcommand{\mco}{\ensuremath{M_\textup{CO-core}}}

\newcommand{\mrem}{\ensuremath{ M_\textup{rem}}}
\newcommand{\mwr}{\ensuremath{ M_\textup{WR}}}

\newcommand{\mbh}{\ensuremath{ M_\textup{BH}}}
\newcommand{\mns}{\ensuremath{ M_\textup{NS}}}

\newcommand{\sevn}{\texttt{SEVN}\xspace}
\newcommand{\parsec}{\texttt{PARSEC}\xspace}
\newcommand{\ace}{\ensuremath{\alpha_{\rm CE}}\xspace}
\newcommand{\kms}{\ensuremath{\rm ~km~s^{-1}}\xspace}

\begin{document} 

  \title{Wolf-Rayet--compact object binaries \\as  progenitors of binary compact objects}
   \titlerunning{Wolf-Rayet--compact object binaries}
  \author{
   Erika Korb\inst{1,2,3}\thanks{\href{mailto:erika.korb@studenti.unipd.it}{erika.korb@studenti.unipd.it}}, 
   Michela Mapelli\inst{3,1,2,4}\thanks{\href{mailto:mapelli@uni-heidelberg.de}{mapelli@uni-heidelberg.de}}, 
   Giuliano Iorio\inst{1,2,4}, 
   Guglielmo Costa\inst{5,4},
   Marco Dall'Amico\inst{1,2,3}
   }
   \authorrunning{E. Korb et al.}
  \institute{
   Physics and Astronomy Department Galileo Galilei, University of Padova, Vicolo dell’Osservatorio 3, I–35122, Padova, Italy \and
   INFN - Padova, Via Marzolo 8, I–35131 Padova, Italy \and
   Institut für Theoretische Astrophysik, ZAH, Universität Heidelberg, Albert-Ueberle-Straße 2, D-69120, Heidelberg, Germany \and
   INAF – Padova, Vicolo dell’Osservatorio 5, I-35122, Padova, Italy \and  
   Univ Lyon, Univ Lyon1, ENS de Lyon, CNRS, Centre de Recherche Astrophysique de Lyon UMR5574, F-69230 Saint-Genis-Laval, France
    }
   \date{Received 11 October, 2024; Accepted, 4 February 2025}

  \abstract{
   Binaries with a Wolf-Rayet star and a compact object (WR--COs), either a black hole (BH) or a neutron star (NS), have been proposed as possible progenitors for the binary compact object mergers (BCOs) observed with gravitational wave (GW) detectors. In this work, we use the open source population synthesis code \sevn to investigate the role of WR--COs as BCO progenitors. We consider an initial population of $5 \times 10^6$ binaries,
   and we evolve it across 96 combinations of metallicities, common envelope efficiencies, core-collapse supernova models, and natal kick distributions. We find that WR--COs are the  progenitors of most BCOs, especially at high and intermediate metallicity. 
   At $Z=0.02,\,{}0.014,$ and 0.0014, more than $\gtrsim 99 \%$ of all the BCOs in our simulations evolved as WR--COs. At $Z = 0.00014$, inefficient binary stripping lowers the fraction of BCOs with WR--CO progenitors to $\approx 83-95 \%$. Despite their key role in BCO production, only $\approx 5-30 \%$ of WR--COs end their life as BCOs. We find that Cyg X-3, the only WR--CO candidate observed in the Milky Way, is a promising BCO progenitor, especially if it hosts a BH. In our simulations, about $70-100 \%$ of the Cyg X-3-like systems in the WR--BH configuration (BH mass $ \leq 10 \msun$) are BCO progenitors, in agreement with the literature. Future observations of WR--COs similar to Cyg X-3 may be the Rosetta stone to interpret the formation  of BCOs.  }

   \maketitle

\section{Introduction}
The LIGO--Virgo--KAGRA (LVK) collaboration has detected almost one hundred gravitational wave (GW) event candidates as a result of its first three observing runs \citep{GWTC-3, Wadekar2023_GWtemplates}. All the confirmed events are associated with binary compact object mergers (BCOs); most of them are binary black holes (BBHs), but there are also binary neutron stars (BNSs; e.g., GW170817 and GW190425; \citealt{GW170817, GW190425}), and black hole-neutron stars (BHNSs; e.g., GW200105 and GW200115; \citealt{GW200105_GW200115_BHNS}). The fourth observing run of LVK is ongoing and has already led to more than a hundred significant GW event candidates.\footnote{Updated list of significant event candidates at \url{https://gracedb.ligo.org/}} Among them, the BHNS system GW230529 comprises a black hole (BH) in the mass range 3--5 M$_\odot$, confirming that there is no empty gap between the maximum neutron star (NS) and the minimum BH mass \citep{GW230529_GWmassgap}.

Binaries hosting a Wolf-Rayet star (WR) and a compact object (CO), either a BH or an NS, are regarded as promising BCO progenitor candidates \citep{Belczynski2012_HMXBHfate}.  Tight WR--CO binaries already host a CO, the WR can end its life as an NS or BH \citep{Limongi2010_preSNevo, parsec2015_chen}, and these systems likely interacted via at least one mass transfer episode - as expected for BNSs, BHNSs, and BBHs progenitors (see, e.g., \citealt{bethe1998,Belczynski2008_STARTRACK,Dominik2012,giacobbomapelli2018_mobse_fryer,VignaGomez2020_CEforBNS,Belczynski2020_popsynthchannel,Bavera2021masstransfer,Santoliquido2021_mergerrate,Broekgaarden2022_formationchannels,Iorio2023_SEVN2,Boesky2024}).

Mass transfer episodes can partially or totally remove the external layers of a star, producing a binary stripped star with a modified internal structure and evolution
\citep[e.g.,][]{Laplace2020_WRradius,Schneider2021_SNfromstrippedstars,Schneider2023_strippedstars}. Stripped stars can form via binary stripping but also via self-stripping, if they have strong stellar winds (typical mass loss rates of WRs are up to $\mdot \approx 10^{-4}$ M$_\odot$; see, e.g., \citealt{KoushikSen2021_WR-OandBH-Obinaries,WRwindstemperaturedependence_Sabhahit2023,WRwindstemperaturedependence_Sander2023,Gotberg23_newWRobservations_2307.00074}). Regardless of the stripping process, stripped stars can be visible as WRs if they are sufficiently hot (temperature at the sonic point $\gtrsim 40\times{}10^3$~K; \citealt{NugisLamers2002_WRwinds,GrafenerVink2013_WRwindsTemperatureRadiusDef,Grafener2017_WRwindssonicpoint}), luminous ($L \gtrsim 10^5$~L$_\odot$ at solar metallicity; \citealt{Shenar2020_WRstarsisolated}), and exhibit emission line-dominated spectra (e.g., helium, carbon, nitrogen lines; see the review by \citealt{Crowther2007_reviewWR}). Observations in the Milky Way (MW), Small Magellanic Cloud, and Large Magellanic Cloud  \citep{VanDerHucht2001_WRcatalogueMW,Bartzakos2001_WRcatalogueMC,Foellmi2003_WRcatalogueSMC,Foellmi2003_WRcatalogueLMC,Schnurr2008_WRcatalogueLMC} indicate that binary stripping is a common formation channel for WRs (binary fractions $\gtrsim 40 \%$), but it does not become dominant at decreasing metallicity, where stellar winds are weaker. \citet{Shenar2020_WRstarsisolated} argue that the WR binary formation channel has a non-trivial dependence on metallicity and is heavily affected by WR spectral definitions and wind models.

Wolf-Rayet--compact object systems are one of the possible electromagnetic progenitors of BCOs. So far, seven WR--CO candidates have been observed (see \citealt{observations} and references therein), but their characterization is challenging. The strong WR winds prevent a precise dynamical CO mass measurement, leaving uncertainties on the BH or NS nature. Uncertainties in the CO nature also affect the only WR--CO candidate known in the MW: Cyg X-3 \citep[e.g.,][]{Cyg-X3_Zd2013, ICX10X-1_Laycock2015_revisited, CygX-3_Koljonen2017, NGC300X-1_Binder2021_BHpreciso}. \citet{Belczynski2013_CygX-3fate} showed that observational and theoretical uncertainties, such as models of core-collapse supernovae (CCSNe) and wind mass loss, significantly affect  the possible fates of WR--COs similar to Cyg X-3 and should be taken into account in evolutionary studies.

In this work, we investigate the role of binaries hosting a WR and a CO as BCO progenitors in the isolated evolution scenario.  We used the population-synthesis code \sevn \citep{Iorio2023_SEVN2} to simulate the evolution of a binary population similar to the MW one under different conditions. We probed a large parameter space to characterize BCO formation and bracket uncertainties in theoretical models. Key parameters for modeling evolutionary processes, such as mass transfer and CCSNe, are still poorly constrained, and their impact on the demography of BCOs is large (see the reviews by \citealt{Mapelli2021_reviewBBHformulacostheta} and \citealt{MandelBroekgaarden2022_modeluncertainties}, and references therein). Here, we systematically explore the role of stellar metallicity, common envelope (CE) ejection efficiency, CCSN physics, and natal kicks. We also discuss the impact of uncertainties in mass transfer physics on the WR--CO and BCO formation channels. Eventually, we consider the implications for BH spins in the case of tidal spin-up.

This paper is organized as follows. In Section \ref{sec:methods} we describe the features of \sevn relevant to this work and the assumptions on the parameter space. In Section \ref{sec:results} we present the results of our simulations and a comparison with Cyg X-3. In Section \ref{sec:discussion} we discuss the impact of mass transfer assumptions, BH tidal spin-up, and other aspects of our parameter-space exploration. In Section~\ref{sec:conclusions} we summarize our results.

\section{Methods}\label{sec:methods}
\subsection{The \sevn code}\label{subsec:sevncode}
We performed our simulations using the \sevn population-synthesis code \citep{Spera2015_remnantspectrum,spera2017_pisnSNe,spera2019_mergingBBH,mapelli2020_compactness,Iorio2023_SEVN2}. \sevn interpolates stellar properties from a grid of pre-computed stellar tracks and implements binary evolution processes through analytic and semi-analytic models. We report hereafter only the assumptions most relevant for this work; we refer to \citet{Iorio2023_SEVN2}  for a more detailed discussion about \sevn.

\subsection{WR definition}\label{subsec:WRdef}
Wolf-Rayet stars are commonly defined based on their spectral characteristics \citep{Crowther2007_reviewWR}, but \sevn does not model the spectral emission of stars directly. Thus, we do not use a spectrum-based definition for WRs but we approximate it with a mass-based definition. We assume that a WR star is a helium star: a core-He or shell-He burning star with helium core mass constituting most of its stellar mass ($\mhecore \geq 97.9 \%~ \mstar$).

From an observational point of view, helium stars are stripped-stars that may or may not already exhibit the spectral features of WRs \citep{Shenar2020_WRstarsisolated}. As the stripping proceeds, the hotter internal layers of the He star are exposed and the surface abundance of advanced burning products is expected to increase. Eventually, the spectrum of a He star will exhibit strong helium, carbon or oxygen emission lines, resembling the one of a WR star of the WN, WC or WO sub-type, respectively \citep{Crowther2007_reviewWR}. Our definition of WR is based on the properties of the WN sub-type at Galactic metallicity and on the WR spectral models elaborated with the \texttt{PoWR} code \citep{HamannGraefener2004_powrWRwinds,Hainich2014_WRobservationPOWRgrids}.

\subsection{Stellar tracks}\label{subsec:tracks}
We interpolate stellar properties from two set of tracks: H-rich stars and pure-He stars. These tracks were calculated with the \parsec stellar evolution code \citep{parsec2012, parsec2015_chen, parsec2019Costa, parsec2022_Nguyenrotation,Costa2025_PARSECtracks}, with the version described in \citet{Costa2025_PARSECtracks}.

H-rich stars evolve from the pre-main sequence (pre-MS) to the end of the carbon burning phase. In such models, the mass-loss rate of hot massive stars depends on  both stellar metallicity and Eddington ratio \citep{VinkdeKoterLamers2000_windsOB,VinkdeKoterLamers2001_windsOB,GrafenerHamann2008_WRwinds,Vink2011}. Specifically, winds of WRs  follow the prescriptions presented by \cite{Sander2019_WRwinds}, with the additional metallicity dependence adopted by \cite{MassGapStellarEvo_Costa2021}.

Pure-He stars evolve from a He zero-age main sequence (ZAMS)  to the end of the carbon burning phase. He-ZAMS stars are initialized accreting mass on a $0.36 \msun$ He core, obtained removing the H-rich envelope from the progenitor star when it started the He-core burning phase. The initial metallicity ($Z$) determines the initial helium mass fraction ($Y=1-Z$) of stars in the He-ZAMS. Winds of pure-He stars follow \citet{Nugis2000_WRwinds}. In this work, we used this set of pure-He tracks to interpolate the properties of the He-stars that we used as proxy for WRs.

Since the definition of WR star depends on the helium core mass, we recall here the most important assumptions for its evolution adopted in \parsec. We define the radius of the He core as the most external shell with hydrogen mass fraction $X < 10^{-3}$. For the convection treatment, we adopt the Schwarzschild criterion \citep{Schwarzschild1958} and the mixing-length theory \citep{Bohm-Vitense1958} with a solar-calibrated value for the parameter $\alpha_{\rm MLT}$ = 1.74 \citep{parsec2012}. We consider a penetrative overshooting scheme above the core boundary limit and we parametrize it with the mean free path $\lambda_{\rm ov}$ of the unstable elements. In the tracks adopted in this work, $\lambda_{\rm ov} = 0.5~ H_{\rm p}$, where $H_{\rm p}$ is the pressure scale height; according to the ballistic approach \citep{Bressan1981_ballisticapproach}. This assumption corresponds to a value of about $f_{\rm ov} = 0.025$ in the exponential decay
overshooting formalism (diffusive scheme; \citealt{Herwig1997_overshooting}). For more details about the \parsec tracks adopted in this work we refer to \citet{Costa2025_PARSECtracks} and \citet{Iorio2023_SEVN2}.

\subsection{CCSN models}\label{subsec:CCSNmodels}
In this work, we tested three CCSN models for CO formation: the rapid and delayed ones proposed by \cite{Fryer2012} and a compactness-based one that uses a compactness parameter \citep[e.g.,][]{Oconnor2011_compactness}, as done in \citet{mapelli2020_compactness}. The three models calculate the remnant mass as a function of the carbon-oxygen core mass at the end of the carbon burning phase (\mco). 

Regardless on the CCSN model, in this work, we used the fiducial \sevn classification for COs \citep{Iorio2023_SEVN2}: we considered a remnant a BH if its mass is $\mrem \geq 3 \msun$, an NS otherwise.

\subsubsection{Rapid and delayed CCSN models}
The rapid and delayed models assume that the shock is revived by the energy accumulated by neutrinos in a convective region between the infalling external layers and the surface of the proto-NS. \cite{Fryer2012} distinguish rapid or delayed explosions considering shocks revived $< 250$ ms or $> 500$ ms after core bounce, respectively. In both models, the remnant mass $M_{\rm rem}$ is expressed as

\begin{equation}
    M_{\rm rem} = M_{\rm proto} +  f_{\rm fb}~ (M_{\rm star} - M_{\rm proto}),
\end{equation}
where $f_{\rm fb}$ (fallback parameter) is the fraction of stellar mass $M_{\rm star}$ that falls back to the proto-compact object ($M_{\rm proto}$). Both $f_{\rm fb}$ and $M_{\rm proto}$ are functions of $\mco$.
The heaviest stars ($\mco \geq 11 \msun$) always undergo a direct collapse, accreting all the ejected material ($f_{\rm fb}=1$), and forming a BH.

\subsubsection{Compactness-based CCSN model}
The compactness-based CCSN model tested in this work uses the compactness parameter $\xi_{2.5}$ \citep{Oconnor2011_compactness} as a proxy for the explodability of stellar cores. The parameter $\xi_{2.5}$ is defined as the ratio between a reference mass ($2.5 \msun$ here) and the radius that encloses such mass at the onset of the core-collapse. In \sevn, we calculate $\xi_{2.5}$ following the fit derived by \citet{mapelli2020_compactness} on non-rotating stellar tracks generated with the \texttt{FRANEC} code \citep{Limongi2018_rotatingCOcompactness}:

\begin{equation}\label{eqn:compactnessDef}
    \xi_{2.5} 
    = \frac{M= 2.5 \msun / \msun}{R(M=2.5 \,{}\textrm{M}_\odot) \ / 1000~\textrm{km}} 
    = 0.55 -1.1 \left(\frac{\mco}{1 \msun} \right)^{-1.0}.
\end{equation}
Several studies suggest that there is a complex relation between the compactness parameter and the carbon-oxygen core mass \citep[e.g.,][]{Sukhbold2018_compactnessnonmonotonic,PattonSukhbold2020_compactnessnonmonotonic,ChieffiLimongi2020_compactnessnonmonotonic,Schneider2021_SNfromstrippedstars,Schneider2023_strippedstars}. In this work, we followed the conservative approach by \citet{mapelli2020_compactness}, based on the results of \citet{Limongi2018_rotatingCOcompactness}, and considered a monotonic relation.

Here, we assumed that if $\xi_{2.5} \leq 0.35$ ($\mco \leq 5.5 \msun$),
the CCSN is successful and produces an NS. The NS mass was drawn from a Gaussian distribution centered at $1.33 \msun$, with standard deviation of $0.09 \msun$ and minimum NS mass set to $1.1 \msun$, in agreement with NSs observations in binaries  \citep{OzelFreire2016_NSmassreviewobservations}. If $\xi_{2.5} > 0.35$,
the star collapses directly to a BH, with mass $M_{\rm BH}$,

\begin{equation}\label{eqn:compactnessMrem}
    \mbh = \mhe + f_{\rm H}~(\mstar - \mhe),
\end{equation}
where $M_{\rm He}$ is the He-core mass at the ignition of the carbon burning and $f_{\rm H}$ is the fraction of the residual hydrogen-rich envelope that collapses  to a BH. The amount of $f_{\rm H}$ is substantially unconstrained.  The work by \cite{fernandez2018} shows that it depends on the binding energy of the envelope at the onset of core collapse. Hence, our choice of $f_{\rm H}=0.9$  should be regarded as a crude approximation.

\subsection{Natal kicks}\label{subsec:kickmodels}
Observations of CO proper motions indicate that compact remnants receive a kick at their birth \citep{Hobbs2005, Verbunt2017_bimodalkicks, Atri2019_kicks}. We tested four models for the magnitude of the kick following the work by \citet{Hobbs2005}, \citet{Atri2019_kicks}, \citet{Fryer2012}, and \citet{SNkicksUnified_Giacobbo2020}. We refer to the corresponding models with the acronyms M265, M70, Mfb, and G20.

\cite{Hobbs2005} fitted the proper motions of young pulsars in the MW with a Maxwellian function with root-mean-square (rms) velocity $\sigma=265 \kms$. Similarly, \cite{Atri2019_kicks} measured the proper motions of BHs in X-ray binaries and fitted their velocity distribution with a Maxwellian with a mean of $107 \pm{} 16$ km s$^{-1}$, corresponding to a rms of $\sigma=70 \kms$. Therefore, in the M265 (M70) model we assign kick magnitudes drawing random numbers from a Maxwellian distribution with rms of $\sigma=265 \kms$ ($\sigma=70 \kms$).

\cite{Fryer2012} assumed that the kick is distributed according to a Maxwellian function with $\sigma=265 \kms$, modulated by the fraction $f_{\rm fb, CCSN}$ of mass that falls back onto the proto-NS:

\begin{equation}\label{eqn:hobbsfb}
    v_{\rm kick} = f_{\sigma=265} ~(1-f_{\rm fb, CCSN}),
\end{equation}
where $f_\sigma{}$ is a random number extracted from the Maxwellian curve. In our simulations with the Mfb model, we choose $f_{\rm fb, CCSN}$ consistently with the CCSN model (see Appendix \ref{app:fallback}).

In the G20 model, we followed \citet{SNkicksUnified_Giacobbo2020}
and re-scaled the kicks drawn from the M265 model:
\begin{equation}
    v_{\rm kick} = f_{\sigma=265} \frac{\langle \mns \rangle}{\mrem} \frac{\mns}{\langle \mej \rangle},
\end{equation}
where $\langle \mns \rangle$ and $\langle \mej \rangle$ are, respectively, the average NS mass and ejecta mass (their values are sensible to the single stellar evolution and CCSN model assumed).

\subsection{Binary evolution and mass transfer}\label{subsec:binaryevo}
Binary evolution processes - such as mass transfer, collisions, and tides - are modeled in \sevn following \citet{Hurley2002}, with a few differences thoroughly described by \citet{Iorio2023_SEVN2}. Hereafter, we summarize the assumptions for mass transfer processes.

A star expanding beyond its Roche lobe donates its mass to the companion in a Roche lobe overflow (RLO) process. As  \citet{Hurley2002}, we determine RLO stability by computing the donor-to-accretor mass ratio $q=M_{\rm d}/M_{\rm a}$ and comparing it to a set of critical thresholds $q_{\rm c}$. If $q>q_{\rm c}$, mass transfer becomes unstable and we either assume that the  two stars merge directly (e.g., if they both fill their Roche lobe), or we start a phase of CE. For a donor star in the main sequence or the Hertzsprung gap phase, we assume $q_{\rm c}=\infty$, since such stars have radiative envelopes, which are less likely to develop unstable mass transfer \citep{Ge2010,Ge2015}. We adopt the same $q_{\rm c}$ values as  \citet{Hurley2002} for stars in more advanced burning stages. We follow their fit on models of condensed polytropes \citep{hjellmingwebbink1987_coreRLOF} for giant stars with deep convective envelopes; we set $q_{\rm c} = 3.0$ for core-Helium burning stars (both H-rich and pure-Helium ones), and we use $q_{\rm c} = 0.784$ for pure-Helium stars with active Helium-shell burning.

We describe the orbital changes due to CE with the usual two parameters $\ace$ and $\lambda_{\rm CE}$ \citep{Webbink1984_CE}. Here, the parameter $\ace$ is the fraction of the orbital energy that is transferred to the envelope, while $\lambda_{\rm CE}$ describes the concentration of the donor's profile. For H-rich stars, we use the fits to $\lambda{}_{\rm CE}$ proposed by \citet{Clayes2014_lambdaCE}, while for pure-He stars we assume $\lambda_{\rm CE}=0.5$ \citep{Iorio2023_SEVN2}. According to the original formalism  \citep{Webbink1984_CE}, the envelope is ejected when the orbital energy lost by the binary equals the envelope binding energy. Thus, \ace cannot be larger than one. However, the orbital energy is not the only source of energy that can be used to eject the CE. Efficient recombination of H and He inside the envelope lowers its binding energy, facilitating its removal. This and other additional sources of energy should be taken into account when modeling CE evolution \citep[e.g.,][]{Paczynski1968,Han1994,Han1995_CErecombinationenergy,Ivanova2013_CE,Klencki2021CE,RoepkeDeMarco2022_CEreview}. Thus, in this work, we also explored values of $\ace > 1$.

\subsection{Simulations and parameter space}\label{subsec:parameterspace}
We wanted to characterize the impact of different physical assumptions on the formation and fate of WR--COs. Thus, we fixed our initial binary population and we evolved it across 96 combinations of parameters and models. We generated a population of $5\times10^6$ massive binary stars with initial mass, mass ratio, orbital period, and eccentricity drawn from observationally motivated distributions \citep{Kroupa2001,Sana2012,MoeDiStefano2017}, as described in Appendix \ref{app:initialcond}. 

We chose four possible values for the metallicity ($Z=$ 0.02, 0.014, 0.0014, 0.00014) and two possible values for the fraction \ace  of the orbital energy required to eject the CE (\ace = 1, 3). For each $(Z,\ace)$ pair, we explored twelve combinations of CCSN and natal kick models: the rapid, delayed, and compactness-based CCSN models, and the M265, M70, Mfb, and G20 natal kick models. In Table \ref{tab:parameterspace} we summarize these combinations and we report the correspondent acronyms.

\begin{table}
    \centering
    \caption{Explored combinations of CCSN and natal kick models.}
    \begin{tabular}{l|cccc}
        \toprule
        \diagbox{CCSN}{Kick} & M265 & M70 & Mfb & G20 \\
        \midrule
        Delayed & dM265 & dM70 & dMfb & dG20 \\
        Rapid & rM265 & rM70 & rMfb & rG20 \\
        Compactness & cM265 & cM70 & cMfb & cG20 \\
        \bottomrule
    \end{tabular}
    \label{tab:parameterspace}
\end{table}

\section{Results}\label{sec:results}
\subsection{Most BCOs form from WR--CO systems} 
\label{subsec:BCOfromWRCO}
\begin{figure*}
    \centering
    \includegraphics[width=\textwidth]{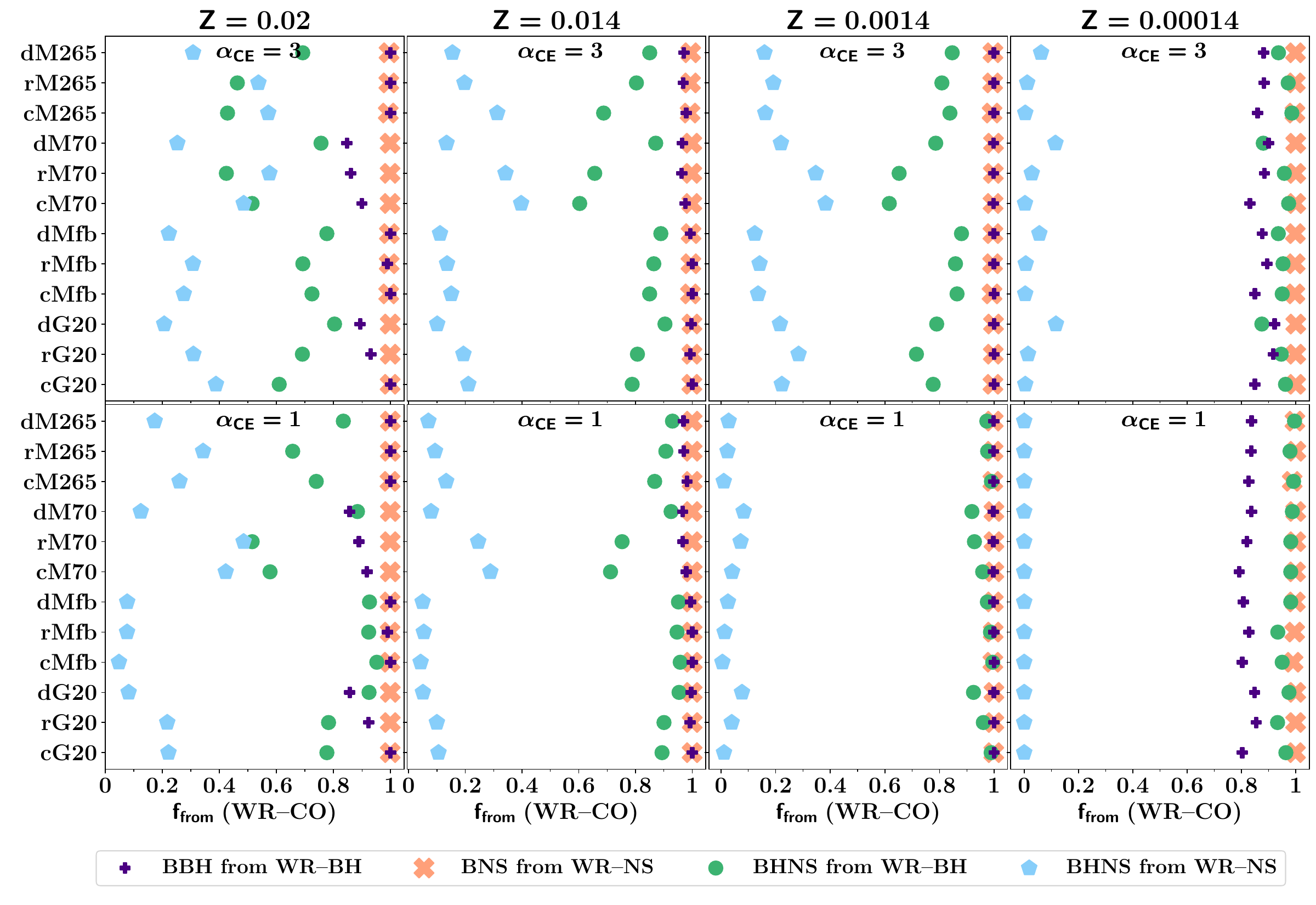}
	\caption{Fraction $f_{\rm from}$ of BCOs with WR--CO progenitors for all the metallicities (columns), $\alpha_{\rm CE}$ efficiencies (rows), and combinations of CCSN and natal kick models tested in this work (horizontal entries, see Table \ref{tab:parameterspace}). We distinguish BCOs into BBHs, BNSs and BHNSs; WR--CO progenitors into systems with a WR star and a BH or NS.}
 \label{fig:GWBCOfrom}
\end{figure*}

For every combination of natal kick, CCSN model, metallicity, and CE efficiency explored in this work (Table \ref{tab:parameterspace}), we find that most BNSs ($\gtrsim$ 99\%), BHNSs ($\gtrsim$ 93\%), and BBHs ($\gtrsim$ 79\%) evolved as WR--BHs or WR--NSs before becoming BCOs (Figure \ref{fig:GWBCOfrom}). In each set, more than $\gtrsim$ 83\% of the total BCO population had a WR--CO progenitor.

\begin{figure*}
    \centering
    \includegraphics[width=\textwidth]{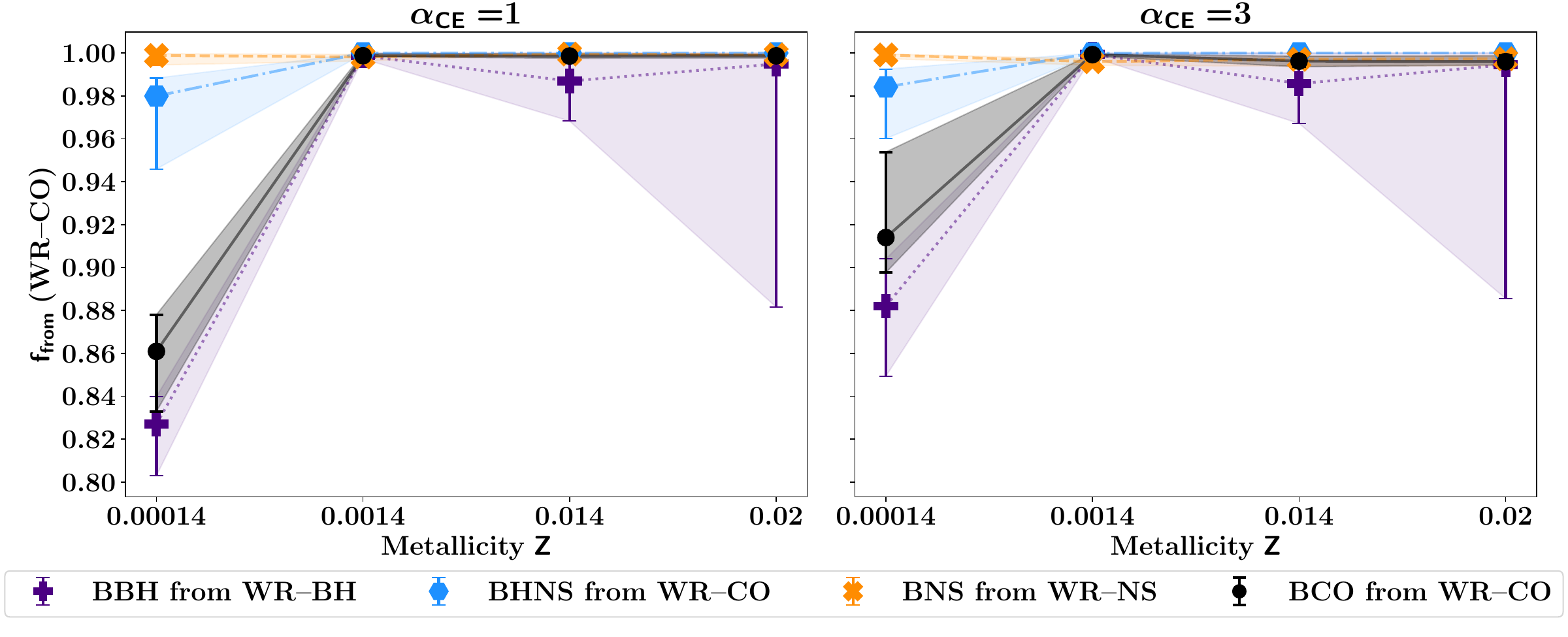}
	\caption{Median fractions \ffrom of BCOs with WR--CO progenitors as a function of metallicity $Z$ and common envelope efficiency \ace. Left (Right) plot shows the $\alpha_{\rm CE}=1$ ($\alpha_{\rm CE}=3$) case. We distinguish fractions \ffrom of BBHs evolved as WR--BHs (purple dotted lines, plus sign markers), BNSs evolved as WR--NSs (orange dashed lines, cross sign markers), BHNSs evolved as either WR--NSs or WR--BHs (blue dash-dotted lines, hexagon markers). We also show the fraction of all BCOs with a WR--CO progenitor (black lines, round markers). For each ($Z, \ace$) pair and BCO type, we consider the distribution of the 12 \ffrom fractions of the correspondent simulated sets (combinations of CCSN and natal kick models, see Tab. \ref{tab:parameterspace}), and we show their median values (scatter points) and 68\% credible intervals (error bars).} 
 \label{fig:GWBCOfromquantiles}
\end{figure*}

Metallicity ($Z$) and CE efficiency ($\ace$) are the quantities that influence more the fraction \ffrom of BCOs with a WR--CO progenitor. Here, we calculated \ffrom as the ratio between the number of BCOs with a WR--CO progenitor (WR--BH or WR--NS) and the total number of BCOs (BBHs, BHNSs or BNSs) in each set. If we fix a pair of ($Z, \ace$) values and consider the distribution of \ffrom across the explored combinations of CCSN and natal kick models, we find that nearly every ($\gtrsim$ 99\%) BCO at $Z\geq 0.0014$ evolved through the WR--CO configuration. At $Z=0.00014$, only $\approx 83-95\%$ of BCOs followed this formation scenario (the range indicates values within the 68\% credible intervals, Figure \ref{fig:GWBCOfromquantiles}). The decrease of the fraction of BCOs with a  WR--CO progenitor at low $Z$ is determined by inefficient stripping of the WR and by the combined effect of CCSN and natal kicks models.

\subsubsection{BHNS progenitors}\label{subsubsec:BHNSprogenitors}
For all the metallicities explored in this work, we find that BHNSs generally form a BH as first CO (Figure \ref{fig:GWBCOfrom}). Evolution through the WR--BH configuration is increasingly favoured for BHNSs at lower metallicities, becoming the dominant ($\gtrsim$ 95\%) formation channel at $Z=0.00014$. For decreasing metallicities, stellar winds are weaker and stars are able to retain more mass \citep{Vink2011}. At $Z=0.00014$, BH production is favoured while NSs require one or more CE events to be produced. Assuming a low CE efficiency ($\ace=1$) completely suppresses the WR--NS channel for BHNSs at $Z=0.00014$.

\begin{figure*}
    \centering
    \includegraphics[width=\textwidth]{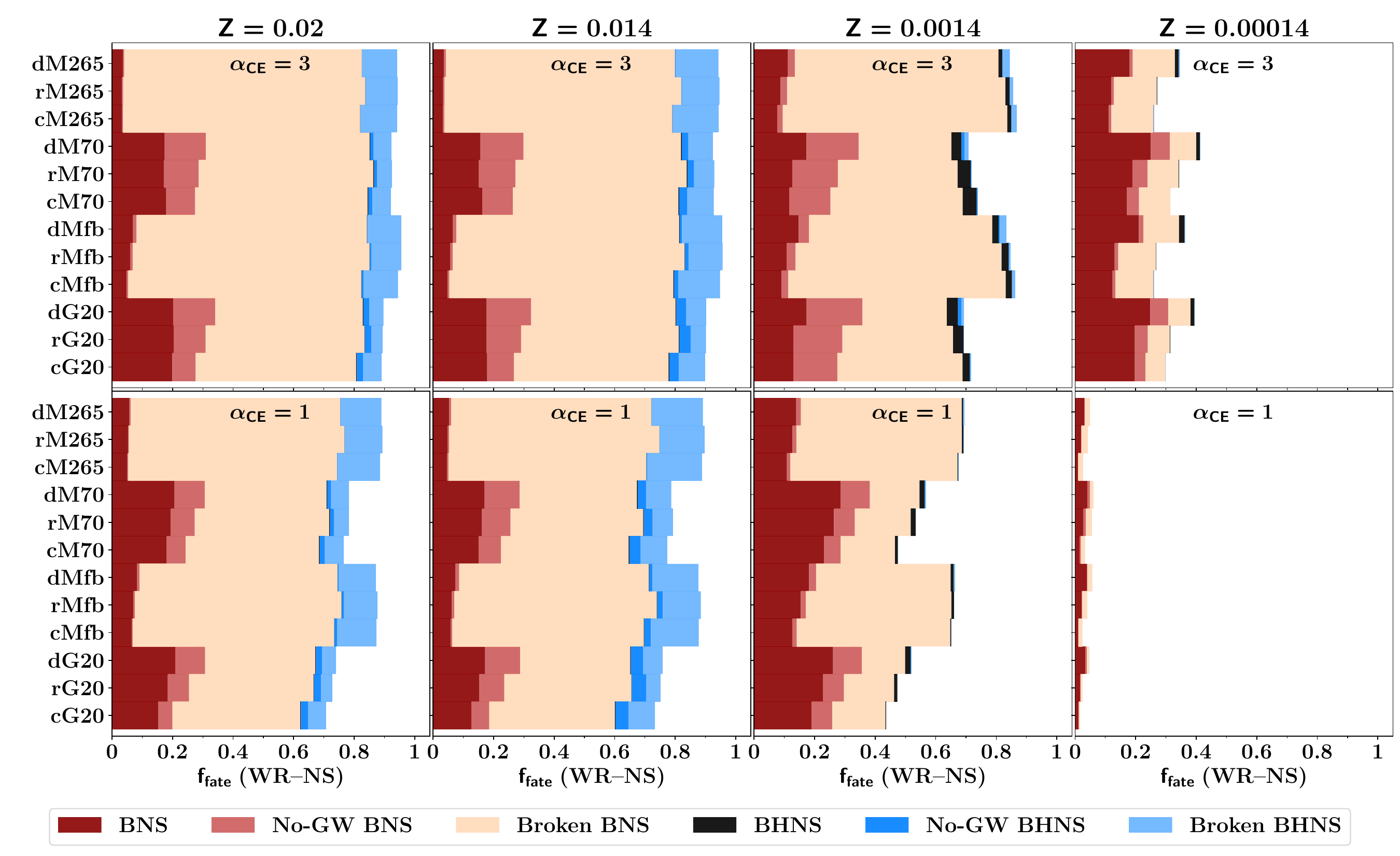}
    \includegraphics[width=\textwidth]{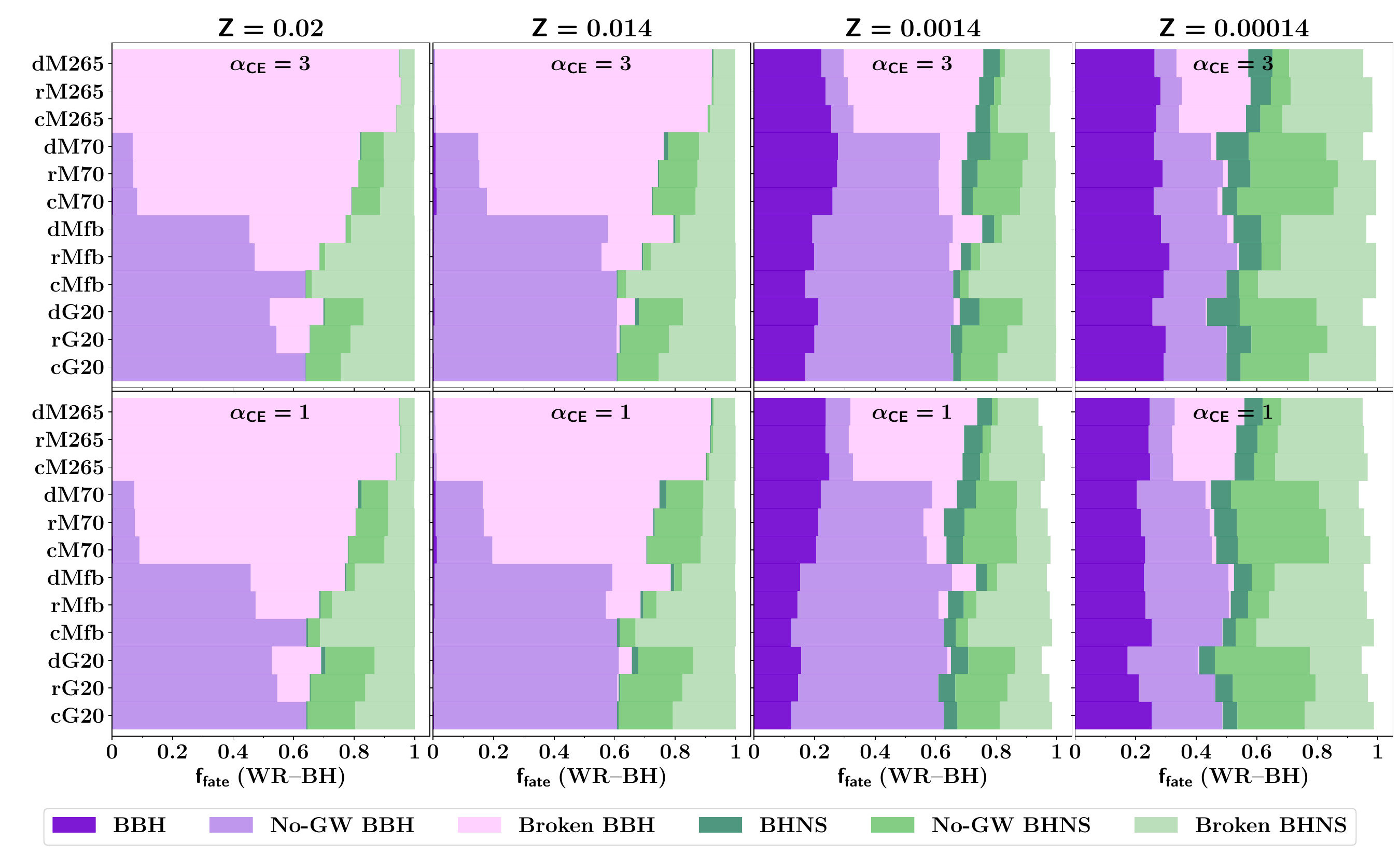}
    \caption{Upper (lower) panels: Fraction \ffate of WR--NSs (WR--BHs) producing a BCO (darkest shades), a bound but not merging BCO (intermediate shades), or resulting in binary ionization (lightest shades). The remaining fraction of WR--NSs (WR--BHs) merges after a collision or a CE episode (not plotted). Each set of panels shows the same fraction \ffate for all the metallicities (columns), $\alpha_{\rm CE}$ (rows), and combinations of CCSN and natal kick models (horizontal entries; see Table \ref{tab:parameterspace}).}\label{fig:WRCOfates}
\end{figure*}

Remnant masses, thus their BH or NS nature, are sensible to the model of CCSN adopted. Assuming different natal kick models influences the natal kick magnitudes and the possibility that a binary is able to survive or merge via GW emission (Figure \ref{fig:WRCOfates}). Thus, the abundance of the WR--BH channel over the WR--NS one is sensible to different combinations of CCSN and natal kick models. For each pair of ($Z,\ace$), we find variations up to $40 \%$ in the fraction of BHNSs that evolved as WR--BHs in place of WR--NSs. These variations are larger at solar metallicity ($Z = 0.02, 0.014$).

\subsubsection{``Missing'' WR--BHs}
For each pair of ($Z,\ace$), binary evolution is determined by the initial orbital properties only until the formation of the first CO. CCSNe impart a natal kick on the newly born CO, modifying the post-CCSN orbital properties, the following evolution of the system, and the possibility that it becomes a BCO and merges within a Hubble time \citep{Peters1964}. As shown in Figure \ref{fig:GWBCOfrom}, the influence of CCSN and natal kicks is usually negligible ($\Delta \ffrom \lesssim 0.01$), except for BHNSs and BBHs at $Z=0.00014$, and BBHs at solar metallicity $Z=0.02,0.014$ (variations can be up to $\Delta \ffrom \lesssim 0.15$). These systems can avoid the WR--CO configuration but still produce BCOs.

Weak stellar winds disfavour the production of WRs via self-stripping, especially at the lowest metallicity explored in this work: $Z=0.00014$. In our simulations, most ($\approx 60-80 \%$) BCOs are BBHs at sub-solar metallicity ($Z=0.0014, 0.00014$). Thus, the lower fraction of BCOs with WR--CO progenitors at $Z=0.00014$ is determined by the lower fraction of BBHs with WR--BH progenitors. 

Binary black holes can form without passing through a WR--BH phase if they experience the second CCSN event while the non-degenerate star is still transferring mass to its BH companion. At low metallicity, if the mass transfer stops before the donor's envelope  is completely stripped, the donor star cannot become a WR  and the system becomes a BBH avoiding the BH-WR stage. We find that these ``missed WRs'' still retain a hydrogen-envelope at the end of their life. Residual envelope masses range from $\approx 50 \%$ to almost $\approx 2\%$ of the total stellar mass, indicating that the most stripped stars were almost WRs according to the definition adopted in this work (He-core mass $\mhecore \geq 97.9 \%~ \mstar$, Section \ref{subsec:WRdef}).

\subsection{Formation channels of BCOs from WR--CO systems}\label{subsec:formationchannels}
\begin{figure*}
	\centering
 \includegraphics[width=.92\textwidth]{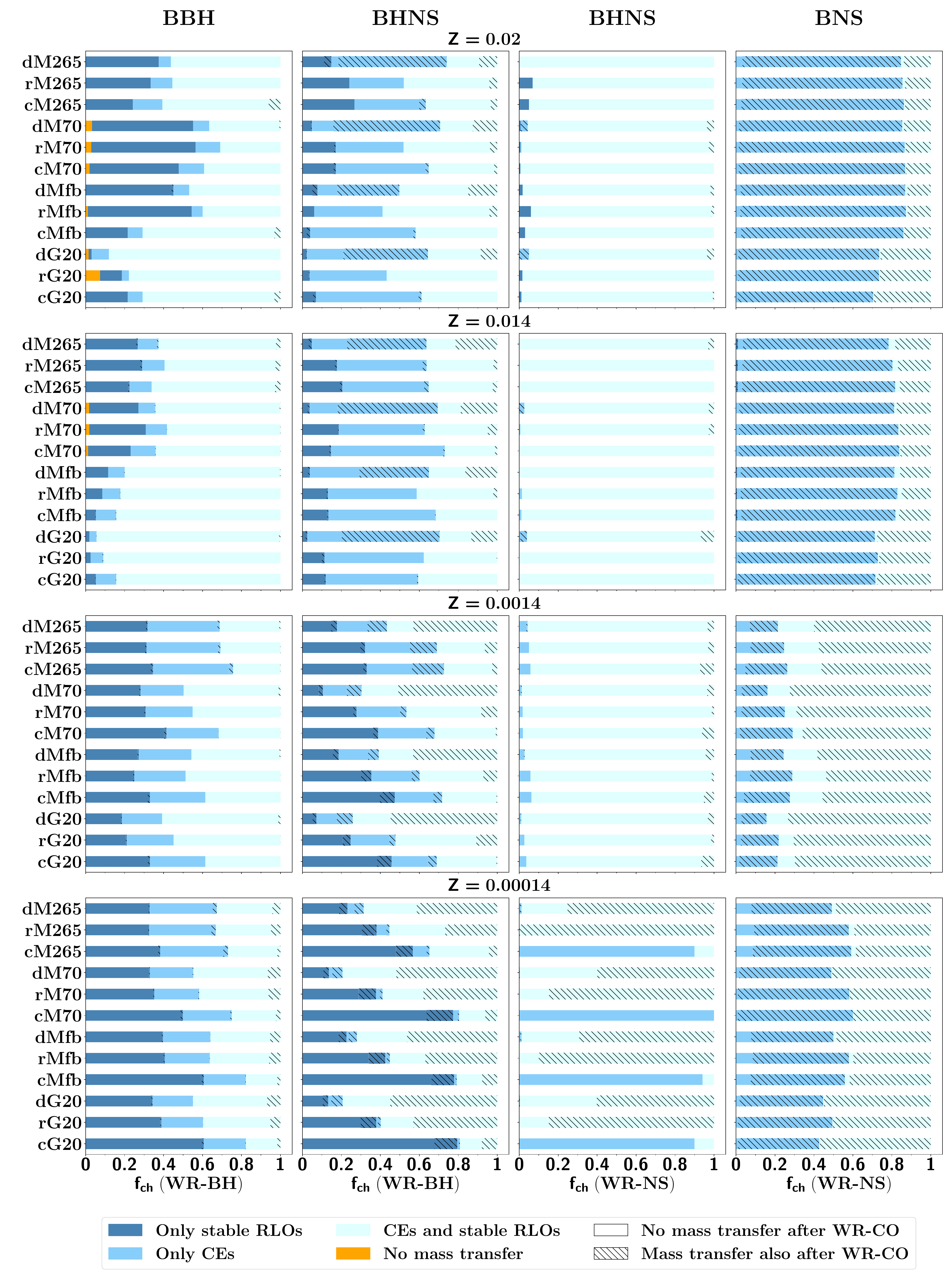}
 \caption{Fraction of BCOs with WR--CO progenitors evolved through different formation channels for  different metallicities (rows) and combinations of CCSN and natal kick models explored in this work (Table \ref{tab:parameterspace}). Here, we show the $\alpha_{\rm CE}=3$ case. We distinguish four main evolutionary paths based on mass transfer events that occurred before the WR--CO formation: stable mass transfer only, CEs only, at least one stable and one unstable mass transfer, no mass transfer. Hatched bars indicate WR--CO systems that become BCOs experiencing at least one additional mass transfer event (stable or unstable RLO) after the WR--CO phase.}\label{fig:channels-a3}
\end{figure*}

When present, the WR--CO configuration constitutes always the last evolutionary stage before BCO formation.  Figure \ref{fig:channels-a3}  shows the formation channels of BCOs with WR--CO progenitors for the $\ace=3$ case (the $\ace=1$ case is shown in the Appendix \ref{appsub:ace1}). We distinguish four channels according to the type of mass transfer events occurring prior to the WR–CO formation: evolution through stable mass transfer only, CE phases only, at least one stable mass transfer and one CE event, no mass transfer at all.

\begin{figure*}
	\centering
 \includegraphics[width=.92\textwidth]{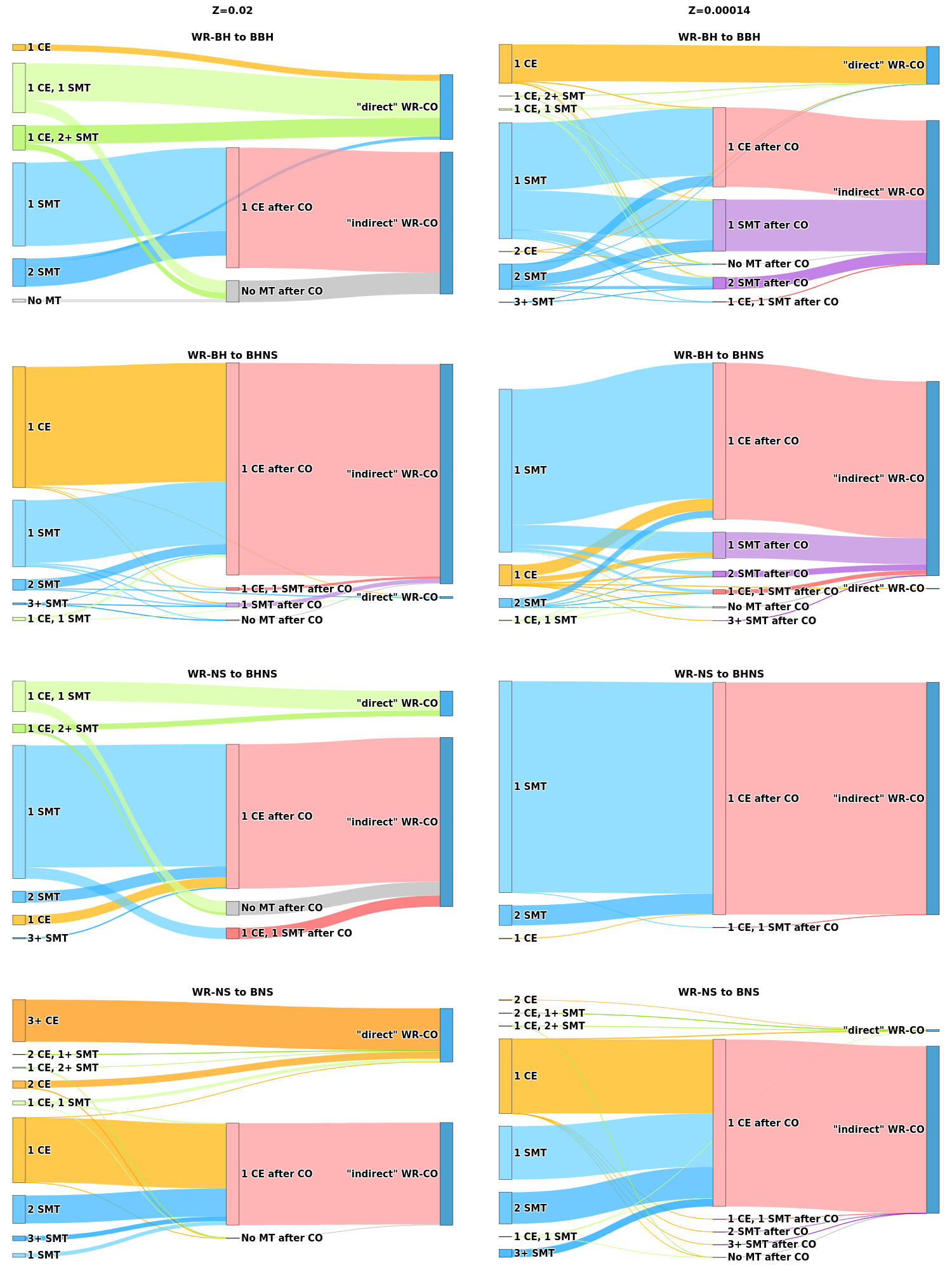}
 \caption{Sankey diagram illustrating formation channels of WR--COs that are also BCO progenitors. We show our fiducial sets, simulated assuming a delayed CCSN model, G20 natal kick model, and \ace=3; for solar ($Z=0.02$, left) and sub-solar ($Z=0.00014$, right) metallicity. We distinguish "direct" WR--COs from "indirect" WR-COs. "Direct" WR--COs are produced by a WR-WR post-CE binary; "indirect" WR-COs avoid the WR-WR evolution. Unlike "direct" WR--COs, WR progenitors in "indirect" WR--COs have the possibilty to start a stable mass transfer (indicated with SMT) or a CE phase between the first CO formation and the WR--CO phase. Few "indirect" systems avoid further mass transfer (indicated with MT) episodes. }\label{fig:sankey}
\end{figure*}

Every family of WR--COs to BCOs (WR--BHs to BBHs, WR--BHs to BHNSs, WR--NSs to BHNSs, and WR--NSs to BNSs) exhibits a preferential formation pathway for a fixed metallicity (Fig.~\ref{fig:channels-a3}). CCSNe, natal kick assumptions, and CE efficiency \ace do not impact the relative fraction of possible formation channels as much as metallicity does. Thus, in Fig.~\ref{fig:sankey} we considered the fiducial sets (simulated with delayed CCSN, G20 natal kick models, and $\ace=3$) to illustrate the main features of the formation channels for these WR--COs at solar ($Z=0.02$) and sub-solar ($Z=0.00014$) metallicity.

WR--COs that are BCO progenitors generally form after one or more mass transfer episodes. Stable mass tranfer episodes or CE events favour the stripping of the WR star and can reduce the orbital separation, facilitating the production of both a WR--CO binary and a system that merges via GW emission within a Hubble time. Only a few ($\lesssim 8\%$) BBH progenitors at solar metallicity form without exchanging mass. This sub-population of non-interacting binaries has wide orbits ($a \gtrsim 10^2~\rsun$) and is able to merge only if the natal kick induced by the second CCSN boosts the eccentricity significantly ($ e \gtrsim 0.9$). Such \emph{lucky kicks} are more frequent for the M70 and G20 natal kicks models at $Z=0.02$.

Stability and duration of the mass transfer phases depend on several factors (e.g., donor mass, mass ratio, stellar radius, stellar phase, orbital separation) and are influenced by metallicity \citep[e.g.,][]{Ge2024}. Thus, as shown in Fig.~\ref{fig:sankey}, there are many possible formation pathways for WR--COs, with large variability in the number and type of stable mass transfer or CE episodes required. However, we can distinguish two main scenarios, corresponding to WR--COs that are produced in a "direct" or "indirect" way. 

In the "direct" scenario, both stars have a similar ZAMS mass ($M_{\rm 2, ZAMS} \gtrsim 0.9 ~M_{\rm 1,ZAMS}$). The primary star initiates a CE event (eventually after other mass transfer episodes) when it is already burning helium in its core, while the secondary is either close to ignite it (end of the red giant branch phase) or is also already burning it. If the CE is successfully ejected, the post-CE configuration consists in two helium cores orbiting about each other: a WR--WR binary has formed. Since the two progenitor stars were already close to the end of their lives, shortly after the end of the CE ($\lesssim 3 \times 10^5$ yr) the initially more massive star (primary at ZAMS) becomes a CO, forming a WR--CO. If the post-CE orbit is tight enough ($a \lesssim 3-4 \,{}\textrm{R}_\odot$), the most massive WR can overfill its Roche lobe and initiate additional short ($\lesssim 5 \times 10^4$ yr) stable mass transfer events before the first CO formation.

In the "indirect" scenario, the WR--CO does not evolve through the WR--WR configuration. The primary star can still experience one or more mass transfer phases as donor star, but at the time of the first CO formation, the secondary companion is not yet a WR. Thus, in this scenario the secondary star can donate mass to the CO before entering  the WR--CO configuration. This scenario is the most common one to produce WR--COs and generally requires the WR progenitor to start a CE episode, in particular at sub-solar metallicity (Figure \ref{fig:sankey}).

Once the WR--CO phase is reached, binaries might undergo wind-fed accretion but they will not experience other mass transfer events to produce BCOs. RLO and CE episodes after the WR--CO phase are common only among WR--COs with light COs, namely WR--NSs or WR--BHs with light BHs, as are the ones produced with the delayed CCSN model ($\sim 3-6 \msun$). Light compact remnants imply long timescales to merge via GW emission \citep{Peters1964}. Thus, to produce a BCO able to merge within a Hubble time, the orbit often has to shrink via mass transfer events: a stable mass transfer phase is generally sufficient for WR--BHs hosting heavier BHs ($ \gtrsim 6 \msun$), especially at lower metallicities (Fig. \ref{fig:channels-a3}), while a CE is preferred if an NS is involved.

\begin{figure*}
        \centering
	\includegraphics[width=\textwidth]{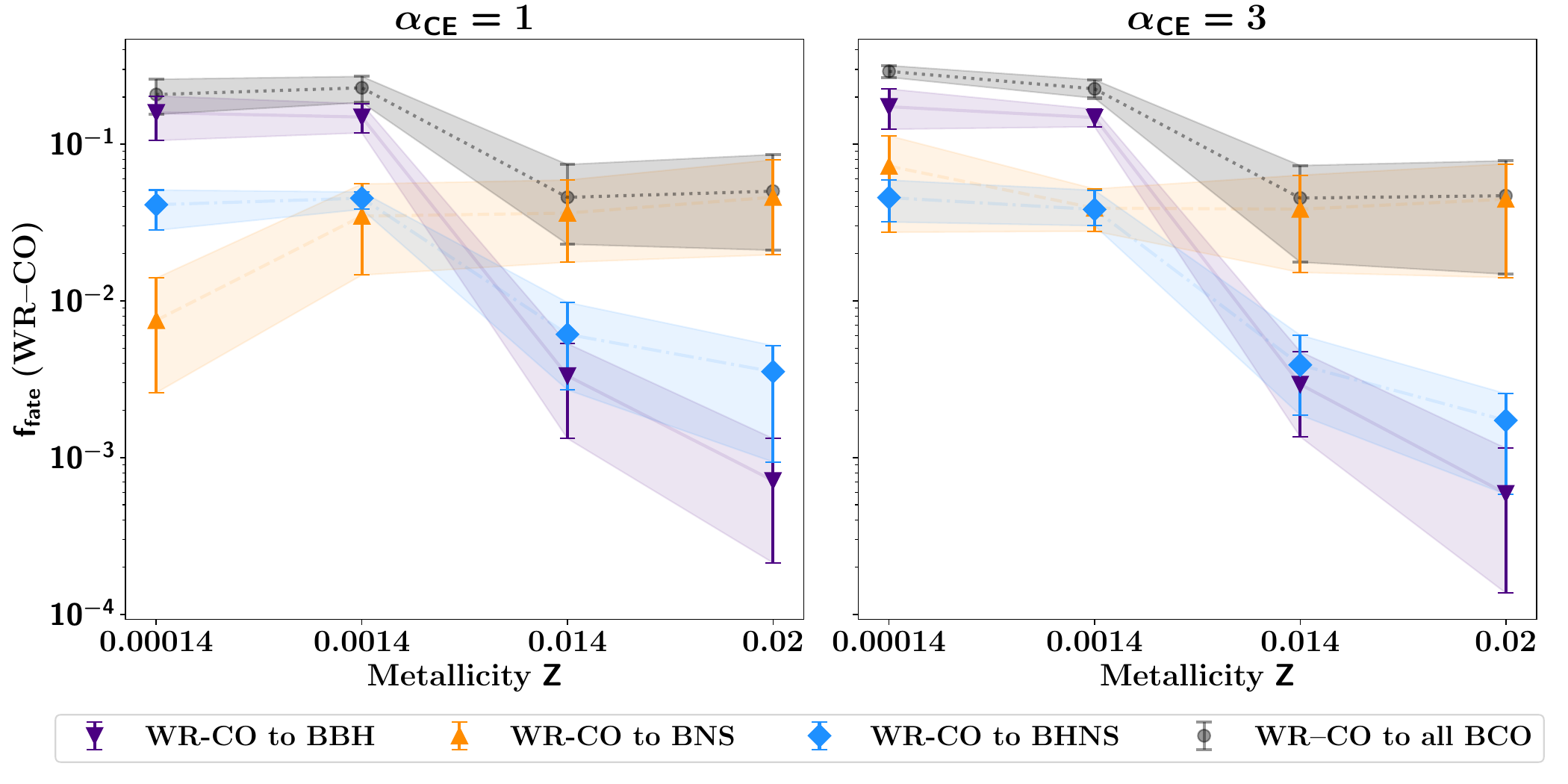}
	\caption{Median fraction \ffate of WR--COs that were able to produce BCOs as a function of metallicity. The left-hand (right-hand) plot refers to $\alpha_{\rm CE}=1$ ($\alpha_{\rm CE}=3$). In each set, we consider all WR--CO systems and we count how many of them form any type of BCOs (gray lines, circle markers), only BBHs (purple lines, lower triangle markers), only BHNSs (blue lines, diamond markers), and only BNSs (yellow lines, upper triangle markers). For each ($Z, \ace$) pair and BCO type, we consider the distribution of the 12 \ffate fractions of the corresponding simulated sets (combinations of CCSN and natal kick models, see Tab. \ref{tab:parameterspace}), and show their median values (scatter points) and 68\% credible intervals (error bars).  }\label{fig:WRCOfatestoGWBCO}
\end{figure*}

\begin{figure*}
	\includegraphics[width=\textwidth]{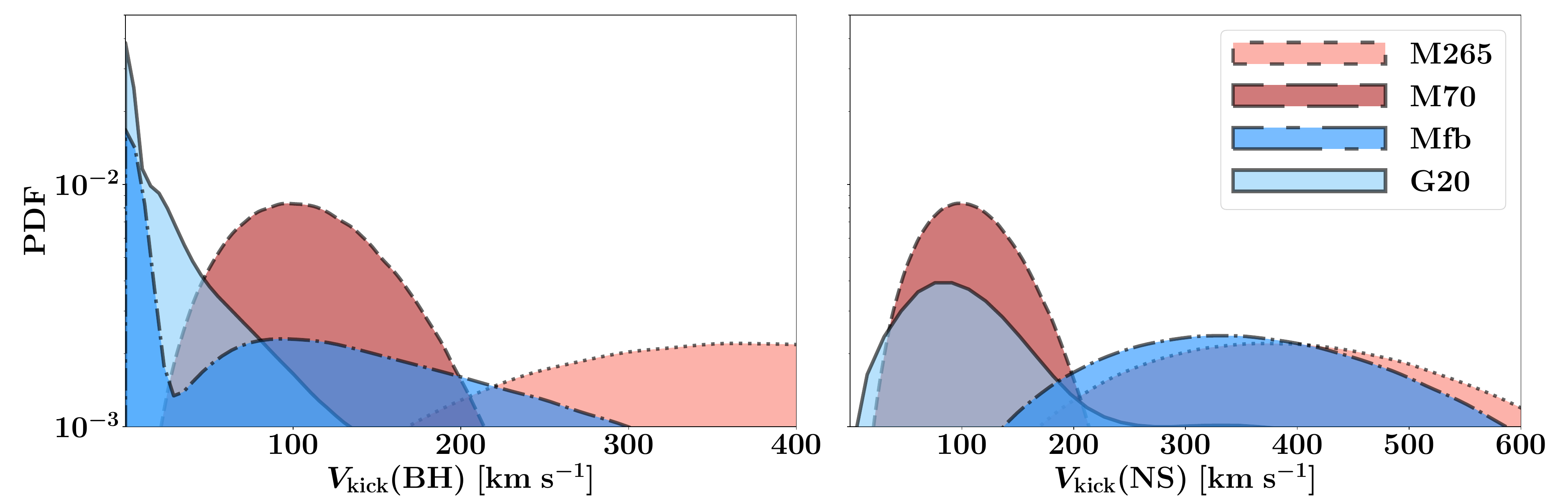}
	\caption{Distribution of natal kick magnitudes received by all BHs (left) and NSs (right) produced by the initial binary population at $Z=0.02$ ($\ace=3$, delayed CCSN) for the four natal kick models explored in this work (Table \ref{tab:parameterspace}).
	}\label{fig:kickdistributions}
\end{figure*}

\subsection{Only a small fraction of WR--CO systems become BCOs} 
\label{subsec:fates}
We showed that the vast majority of BCOs evolved via a WR--CO stage. However, only few WR--COs end their life as BCOs. At sub-solar metallicity ($Z = 0.0014$, $Z=0.00014$), about $\approx 30\%$ of WR--CO systems produce BCOs; at solar metallicity ($Z=0.02$, $Z=0.014$) only $\approx 5\%$ (Figure \ref{fig:WRCOfatestoGWBCO}).

WR--COs are not common: in each simulated set, we find that only $\approx 1-5 \%$ of the initial binary population evolves as a WR--CO system. In our simulations, the relative abundance of WR--BHs (WR--NSs) in the WR--CO population is not directly correlated with the abundance of BBHs (BNSs) in the BCO population. WR--BHs constitute always $\approx 50-80 \%$ of the WR--CO population, with the relative abundance to WR--NSs that is sensible to the CCSN and natal kick models. In contrast, the relative abundance of BBHs in the BCO population is a strong function of metallicity. For instance, BBHs constitute the dominant BCO family at sub-solar metallicity ($\approx 60-80 \%$) but are rare at solar metallicity ($\lesssim 10 \%$). Similar arguments apply to the relative abundance of WR--NSs and BNSs, even if BNS production is also sensible to \ace.

To investigate what determines the fate of WR--COs, we considered the fraction \ffate of WR--CO systems with a specific fate. For instance, to calculate the fraction \ffate of WR--COs becoming BCOs, we consider the number of BCOs with a WR--CO progenitor (WR--BHs, WR--NSs, or WR--BH plus WR--NSs; for BBHs, BNSs or BHNSs, respectively) and we divide it for the correspondent total WR--CO population (all WR--BHs, WR--NSs or WR--COs, respectively). WR--COs could have various possible fates: the system could prematurely coalesce (as a result of a collision or CE event); the WR could form the second CO after the binary split; the binary could be loose and not merging via GWs; the binary could be tight enough to merge via GW emission within a Hubble time. In Figure \ref{fig:WRCOfates}, we show the fractions \ffate calculated for WR--NSs and WR--BHs for all the possible fates and across the parameter space explored in this work (Table \ref{tab:parameterspace}). In Figure \ref{fig:WRCOfatestoGWBCO}, we show median values and $68 \%$ credible intervals for the distributions of \ffate leading only to BCO production for a fixed pair of ($Z,\ace$) values.

\subsubsection{WR--COs and \emph{lucky kicks}}\label{subsubsec:luckykicks}
As shown in Figure \ref{fig:WRCOfates}, most of the WR--NSs that are able to survive in bound orbits have also a final orbital configuration that is optimized to merge via GW emission within a Hubble time. These systems were selected by \emph{lucky kicks}: natal kicks that are strong enough to result in highly eccentric orbits but not too strong to ionize the binary.

Figure \ref{fig:kickdistributions} shows that the G20 and M70 models have natal kick magnitude distributions that peak in the optimal velocity regime of \emph{lucky kicks} for NSs ($ v \approx 100$ \kms). Since BHs are heavier than NSs, the optimal regime for \emph{lucky kicks} of BHs is shifted to higher kick velocities ($v \approx 200 \kms$), where none of the natal kick distributions considered in this work peaks. Therefore, in our simulations, \emph{lucky kicks} boost only the production efficiency of the lightest second-born COs, explaining the higher production efficiency of BNSs and BHNSs with respect to BBHs at $Z=0.02$ (Figure \ref{fig:WRCOfatestoGWBCO}). 

We highlight that selection effects due to \emph{lucky kicks} are responsible also for the higher fraction of BBHs without a WR--BH progenitor at $Z=0.02$ (Figure \ref{fig:GWBCOfrom}). In this case, the selection effect is relevant only for the population of lightest BHs, especially for BHs with mass $\lesssim 6 \msun$ produced by the delayed CCSN model. For decreasing metallicity, the BH mass increases and the effect of \emph{lucky kicks} is negligible. 

\subsubsection{WR--NSs and CE evolution}\label{subsubsec:WRNS suppressed}
CE evolution is the most common channel to form binaries with one or two NSs (WR--NSs, BHNSs, BNSs) and is necessary to produce BNSs at every metallicity (Figure \ref{fig:channels-a3}). Unstable mass transfer events may completely remove the stellar envelope and facilitate the creation of NSs. However, entering a CE phase may cause a reduction of the semi-major by $\Delta a \approx 2000 \rsun$ and a premature collision. This possibility is enhanced for low \ace~ values. The sets evolved with \ace=1 need to extract more orbital energy to expel the CE with respect to sets evolved with \ace=3, often leaving no surviving system.

Collisions driven by CE events reduce the fraction of WR--NSs that can successfully form BHNSs or BNSs. At the lowest metallicity ($Z=0.00014$), NSs are produced via binary stripping and the effect of inefficient CE ejection is more evident. At \ace=1, $\gtrsim$ 90\% WR--NSs coalesce prematurely, limiting the production of BNSs (Figure \ref{fig:WRCOfatestoGWBCO}) and suppressing the formation of BHNSs from WR--NSs (Figure \ref{fig:GWBCOfrom}).


\begin{figure*}
        \centering
        \includegraphics[width=\textwidth]{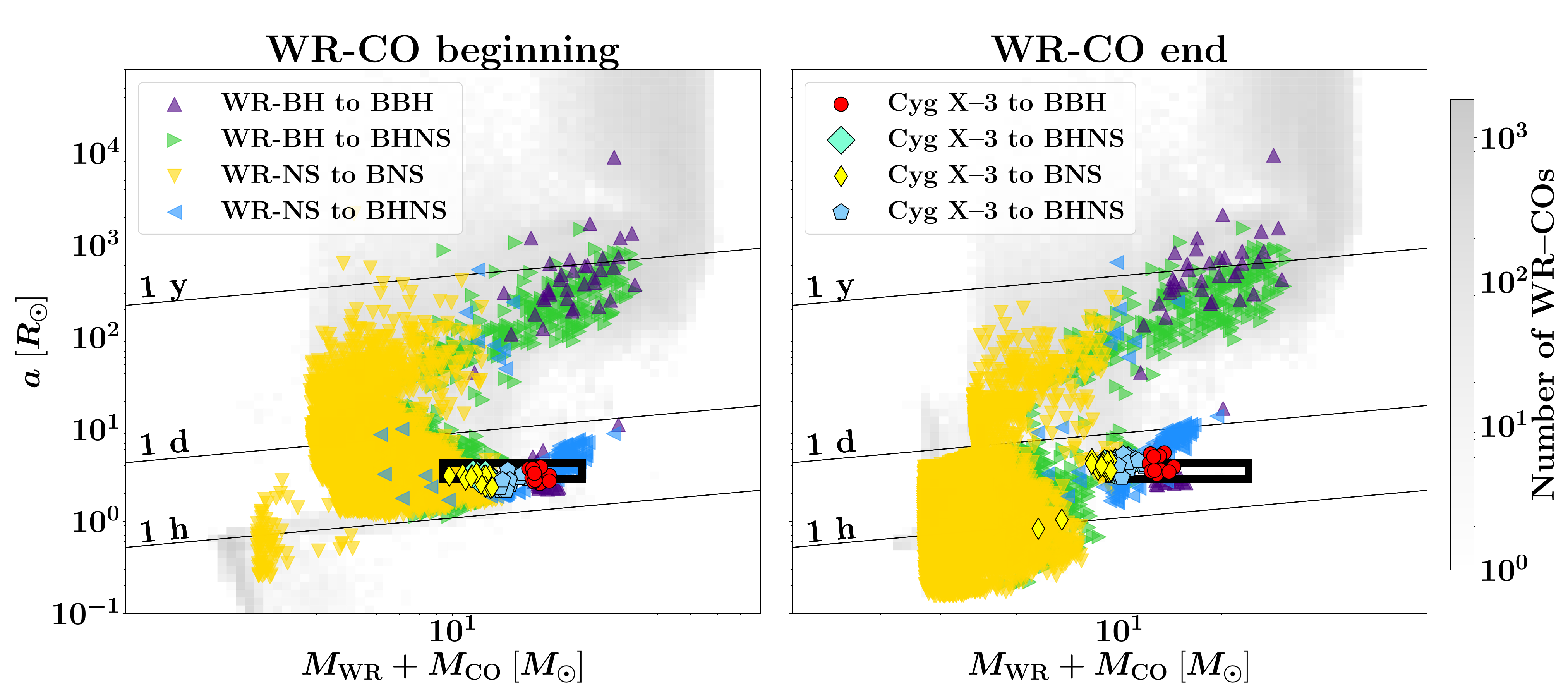}
	\caption{Orbital properties of WR--COs and Cyg X-3 at the beginning (left) or end (right) of the WR--CO configuration. We highlight the properties of WR--COs (triangular markers) and Cyg X-3-like binaries (non-triangular markers) that are BCO progenitors, comparing them to the properties of the whole sample of WR--CO (gray areas in the background). We distinguish binaries ending up as BBHs (purple upward triangles; red circles), as BNSs (gold downward triangles; yellow thin diamonds), or as BHNSs, either forming the BH first (green rightward triangles; thick turquoise diamonds) or the NS (blue leftward triangles; turquoise pentagons). The black rectangular box indicates the region of orbital parameters that we use to select Cyg X-3 candidates (Section \ref{subsec:CygX3}). Black diagonal lines indicate orbital periods of 1 hour, 1 day or 1 year. Here we show the set evolved with $\ace=3$, $Z=0.02$, G20 natal kick model and delayed CCSN model. In Appendix \ref{appsub:CygX3} we show other sets at $\ace=3$ and $Z=0.02$.}\label{fig:CygX3orbital}
\end{figure*}

\begin{figure*}
        \centering
        \includegraphics[width=0.49\textwidth]{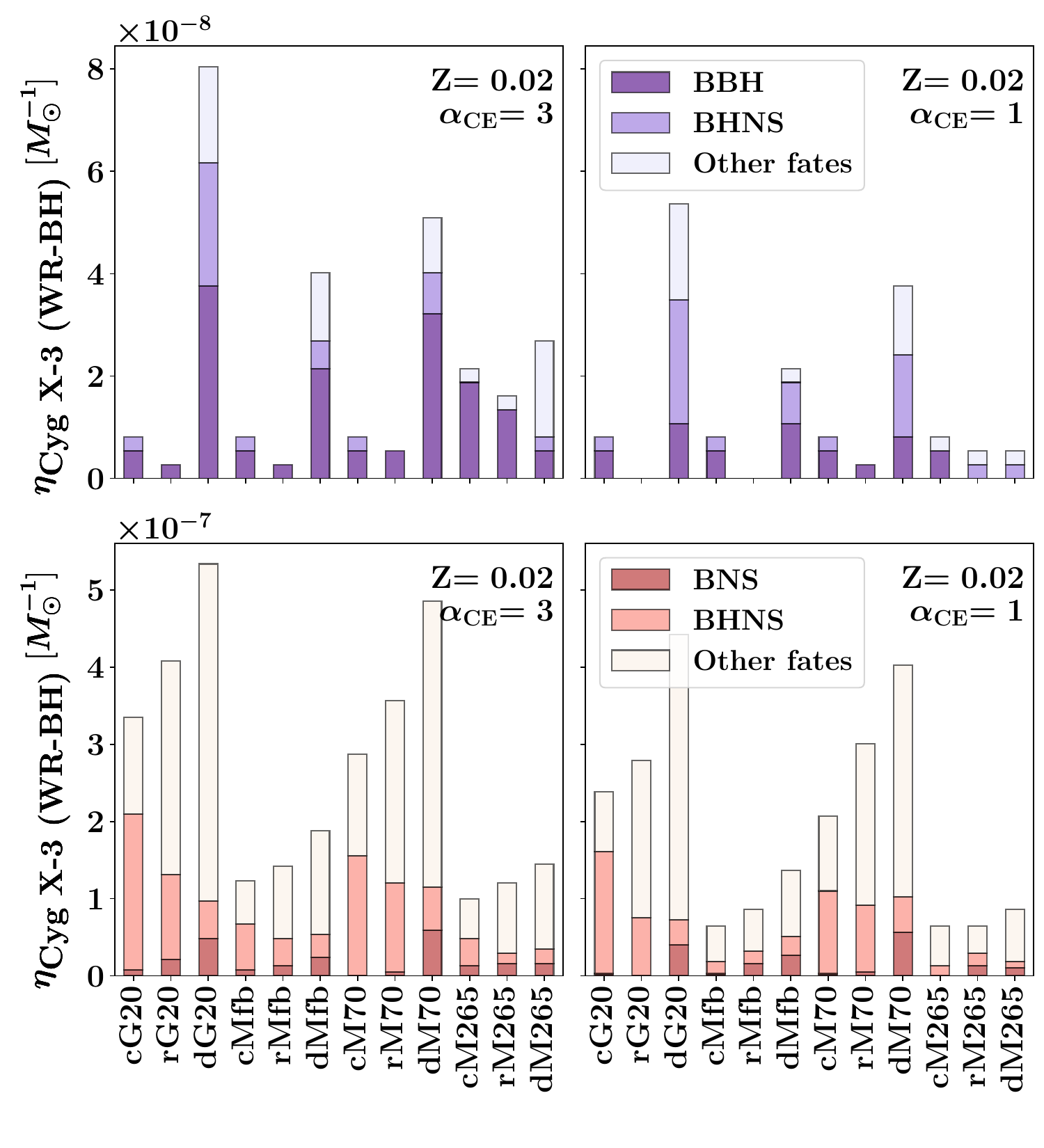}
        \hfill 
	\includegraphics[width=0.49\textwidth]{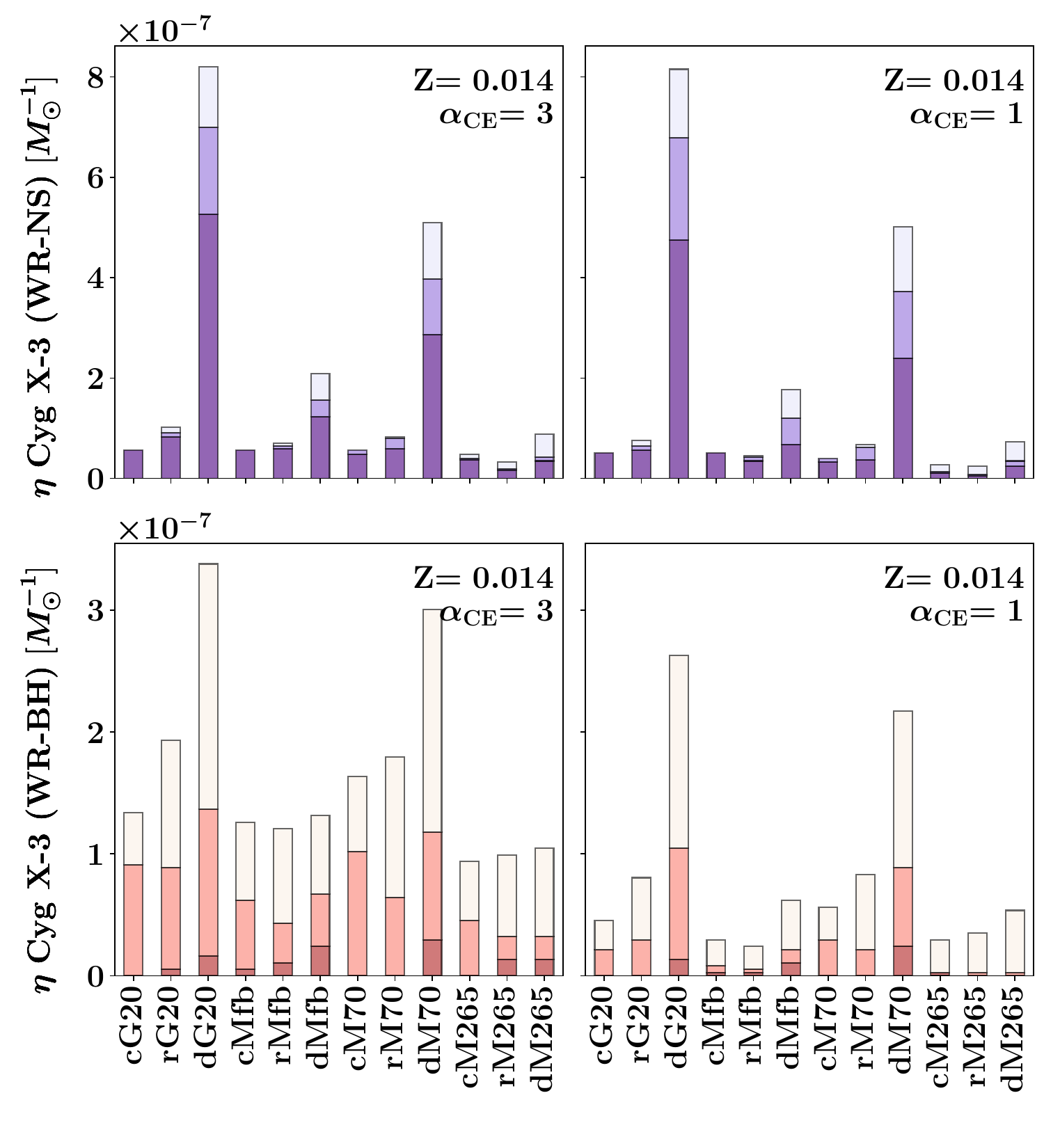}
	\caption{Formation efficiency of Cyg X-3 candidates \citep{CygX-3_Singh2002, CygX-3_Koljonen2017} for each set of binaries simulated at the two solar metallicity values considered in this work: $Z=0.02$ (left) and $Z=0.014$ (right). We distinguish Cyg X-3 candidates hosting a BH (upper panels) or an NS (lower panels). We highlight with darker shades the Cyg X-3 candidates that are BCO progenitors.}\label{fig:NCygX3}
\end{figure*}

\subsection{Cyg X-3: A possible BCO progenitor}\label{subsec:CygX3}
Cyg X-3 is the only WR--CO candidate known in the MW \citep{Cyg-X3_Zd2013,observations, Veledina24_CygX3reviewULX} and has been indicated as a possible BCO progenitor by \citet{Belczynski2013_CygX-3fate}. Here, we investigate this possibility, looking for Cyg X-3--like binaries in our simulations and analyzing their fates. In this work, we define a Cyg X-3--like system as a WR--CO system if its orbital properties satisfy three criteria: orbital period $P_{\rm orb} \in [4.5,5.1]~ \rm hrs$, WR mass $\mwr \in [8-14] \msun$, and CO mass $\mrem \leq 10 \msun$ \citep{CygX-3_Singh2002,CygX-3_Koljonen2017}. We take a conservative approach on the mass and nature of the CO hosted in Cyg X-3 and we allow for both the BH ($M_{\rm rem} = 3-10~\rm \msun$) and NS hypothesis ($M_{\rm rem} < 3~\rm \msun$), given the typical uncertainties in the CO mass determination of COs with WR companions \citep{ICX10X-1_Laycock2015_revisited, NGC300X-1_Binder2021_BHpreciso}.

Cyg X-3 lies in the Galactic plane at a distance of $d_{\rm Cyg X-3 }=7.41 \pm 1.13 ~\rm kpc$ from the Sun \citep{CygX-3_McCollough2016_Observation}. According to the metallicity gradient of the MW derived from Gaia DR2 \citep{MetallicityGradientMW2018Gaia}, Cyg X-3 is expected to have solar-like metallicity ([Fe/H] $\approx$ -0.088). Given the uncertainties in the metallicity determination, we investigated both solar values considered in this work: $Z=0.02$ and $0.014$.

\subsubsection{Cyg X-3: Optimal properties for BCO formation}\label{subsubsec:CygX3orbital}
We find that only $\approx 0.1-0.01 \%$ of WR--CO binaries at solar metallicity have a Cyg X-3-like configuration. Typical WR--CO separations at solar metallicity in our simulations are of $a \gtrsim 10^2 \rm~ \rsun$ (Figure \ref{fig:CygX3orbital}). Such large orbital separations usually require \emph{lucky kicks} in the second-CO formation to increase the eccentricity and allow for BCO production. In contrast, Cyg X-3 has a semi-major axis that is  at least two order of magnitudes smaller ($a \approx 3-5 \rm ~ \rsun$), increasing the possibility that the system will merge via GW emission within a Hubble time. Figure \ref{fig:CygX3orbital} shows that Cyg X-3 occupies one of the regions of the space of orbital properties that strongly favors BCO production. 

About $\approx 20-80 \%$ of the Cyg X-3 candidates in our simulations are BCO progenitors. More than $\gtrsim$ 70\% ($\approx 100 \%$ for some combinations of the parameter space, Figure \ref{fig:NCygX3}) of the Cyg X-3-like binaries hosting a BH end their life as BCOs. Even if we consider evolution with the M265 natal kick model, that generates high-velocity kicks typical of Galactic pulsars \citep{Hobbs2005}, we find that more than $\gtrsim$ 30\% of Cyg X-3 candidates hosting a BH  become BCOs. In contrast, only $\lesssim 60 \%$ of our Cyg X-3-like systems  composed of a WR and an NS become BCOs, because most of the systems break during the second CCSN. Thus, we find that Cyg X-3 has a higher probability to become a BCO progenitor if it hosts a BH, in agreement with \citet{Belczynski2013_CygX-3fate}.

\subsubsection{Formation efficiency and evolution}
Figure \ref{fig:NCygX3} shows the formation efficiency $\eta_{\rm Cyg X-3}$ of Cyg X-3 for all the sets evolved at solar metallicity. For each set, we calculated $\eta_{\rm Cyg X-3}$ as the ratio between the number of Cyg X-3 candidates and the total simulated mass, correcting for the incomplete initial mass sampling of the primary (Appendix \ref{app:initialcond}) and assuming an initial binary fraction of 1 for simplicity. 

The delayed CCSN model is the only CCSN model explored in this work that produces COs in the $\approx 2-6 \msun$ range. Therefore, the delayed model produces more Cyg X-3-like binaries ($M_{\rm rem} \leq 10 \msun$) with respect to the rapid and compactness-based models. Assuming the delayed CCSN model, the formation efficiency of Cyg X-3 can be up to one order of magnitude higher with respect to values obtained in other sets, reaching values of $\eta_{\rm Cyg X-3} \approx 5.6 \times 10^{-7} ~ \msun^{-1}$.

We find that Cyg X-3-like binaries evolve similarly to WR--COs at the same metallicity (Section \ref{subsec:fates}). Cyg X-3 progenitors experience one or two mass-transfer events (with preference for two, one stable and one unstable), but there are no further mass transfer episodes after the system becomes a  WR--CO, except for some WR--NSs. At the beginning of the Cyg X-3-like phase, the WR star does not fill more than $\lesssim 80 \%$ of its Roche lobe: the binary can be wind-fed and the orbital separation can increase, as it is visible in Figure \ref{fig:CygX3orbital}. The second compact object forms after less than $\lesssim 0.15$ Myrs of evolution in the Cyg X-3 configuration.

The G20 and M70 natal kick models select a larger number of Cyg X-3 candidates, especially if they host an NS (Figure \ref{fig:NCygX3}). These binaries entered in the Cyg X-3 region of the parameter space as a consequence of the natal kick. Most of these systems will break after the second CCSN or coalesce prematurely, because of the additional mass transfer events that WR--NSs may experience.

\section{Discussion}\label{sec:discussion}
\subsection{WR formation}\label{subsec:disc-winds}
In this work, WR stars can form either via self-stripping, as a result of wind mass loss, or via binary stripping, as a result of mass transfer events. WR formation via self-stripping requires powerful stellar winds. Since wind mass loss rates depend on the mass of the star and its metallicity 
\citep{Vink2001,Vink2011,Nugis2000_WRwinds,WRwindstemperaturedependence_Sander2023}, in our simulations self-stripping can produce WRs at solar metallicity ($Z=0.02, 0.014$) but not at sub-solar metallicity ($Z=0.0014, 0.00014$). For instance, according to the \parsec tracks adopted in this work, a star evolved in isolation $Z=0.02$ becomes a WR ($\mhecore \geq 97.9 \%~ \mstar$) if it has an initial mass $\mzams \gtrsim{} 30 \msun $ (WR mass of $\mwr \approx 13 \msun$). The same $\mzams \approx 30 \msun $ star, if evolved at $Z=0.0014$,  will still retain $\approx 50\%$ of its mass in its hydrogen-envelope at the end of the carbon burning phase.

We expect that changes in the input physics, in the internal mixing properties, and in the choice of the stellar evolution code adopted would modify mass losses, the internal structure, and the core growth of the star, influencing the thresholds for WR formation via single stellar evolution (e.g., \citealt{Poojan2022_stellarevodifferences} and references therein). For instance, \citep{DallAmico2025} showed that accretion-induced chemically quasi-homogeneous evolution can enhance the formation of WR stars over red supergiants in secondary stars, increasing the fraction of binaries that are able to form WR--CO systems.

\subsection{Mass transfer uncertainties}
In \texttt{SEVN}, we use critical donor-to-accretor mass ratios $q_{\rm c} = M_{\rm donor} / M_{\rm accretor}$ to determine mass transfer stability, according to the formalism adopted in \texttt{BSE} \citep{Hurley2002} and other population-synthesis codes. 
Several works in the literature highlight that this is a simplistic approach to a complicated problem, and has to be revised. For instance, the works by \citet{Ge2010}, \citet{Ge2015}, \citet{Ge2020}, and \citet{Ge2024} indicate that there is no universal $q_{\rm c}$ threshold for a fixed evolutionary phase of the donor star; rather, their values are strongly influenced by assumptions on the stellar physics, metallicity, and conservation of mass and angular momentum. \citet{Willcox2023_MTnonconservativeCOMPAS} find that prescriptions calculated assuming conservative mass transfer could underestimate the number of mergers produced by unstable mass transfer. \citet{gallegos2021MESAvspopsynth} showed that setting a $q_{\rm c}$ threshold favors CE evolution as the dominant formation channel for BBHs, while more detailed stellar evolution calculations favour stable mass transfer evolution as the dominant channel. The work of \citet{gallegos2021MESAvspopsynth} indicates that this approach to mass transfer processes in population synthesis codes could overestimate the merger rate density of BBHs
\citep[e.g.,][]{Neijssel2019,VanSon2022_stableMTexplainsGWmass,Olejak2024_BBHspinsMTstable,Picco2024}. 

In this work, we did not put a $q_{\rm c}$ threshold for MS and HG donor stars: we assumed that they experience only stable mass transfer events, avoiding CEs. This setup corresponds to the fiducial model presented by \citet{Iorio2023_SEVN2}. The resulting BCO formation efficiency and channels are  compared with the ones obtained assuming, for instance, the $q_{\rm c}$ setup of \citet{Hurley2002}. From Figs. 14 to 20 of \citet{Iorio2023_SEVN2} we can see that forcing mass transfer to be stable for MS and HG donor stars has minor effects on BBH properties, while it increases BNS merger efficiency at solar metallicity (by about one order of magnitude).

\subsection{Spin distributions in BBHs}\label{subsec:disc-spins}
\subsubsection{Spin magnitude of second-born BH}\label{subsubsec:disc-spinmagnitude}
According to our models, most BBHs emerge from a WR--BH configuration. The first-born BH can be non-spinning ($\chi_{\rm BH,1} \approx 0$) if the progenitor star loses angular momentum efficiently. In close orbits, as are the ones leading to BBHs, the BH can tidal spin up the WR star. After the CCSN, the second-born BH could be highly spinning and become the dominant contribution to the effective and precessing spin parameters ($\chi_{\rm eff}$ and $\chi_{\rm p}$) that we can measure from GW observations \citep{GWTC-3_interpretation}. 

\citet{Bavera2021_BHspinfromWRtheory} used \texttt{MESA} \citep{MESA2015BinaryMT} to follow the tidal spin up in WR--BH binaries. They assumed efficient angular transport, resulting in non-spinning first-born BH. 
\citet{Bavera2021_BHspinfromWRtheory} derived a fitting formula to predict the dimensionless spin magnitude of the second-born BH $\chi_{\rm BH,2}$ as a function $f$ the orbital period of the WR--BH binary $P_{\rm WR-BH}$ and the WR mass, \mwr, at He or C depletion \citep{Bavera2021_BHspinfromWR}:

\begin{equation}
\chi_\mathrm{BH,2} =  
\begin{cases}
f (P_{\rm WR-BH}, M_{\rm WR}) & P_{\rm WR-BH} \leq \text{ 1 day} \\
0 & \text{otherwise}
\end{cases}.
\end{equation}\label{eqn:baveraspin}

\begin{figure*}
        \centering
        \includegraphics[width=0.328\textwidth]{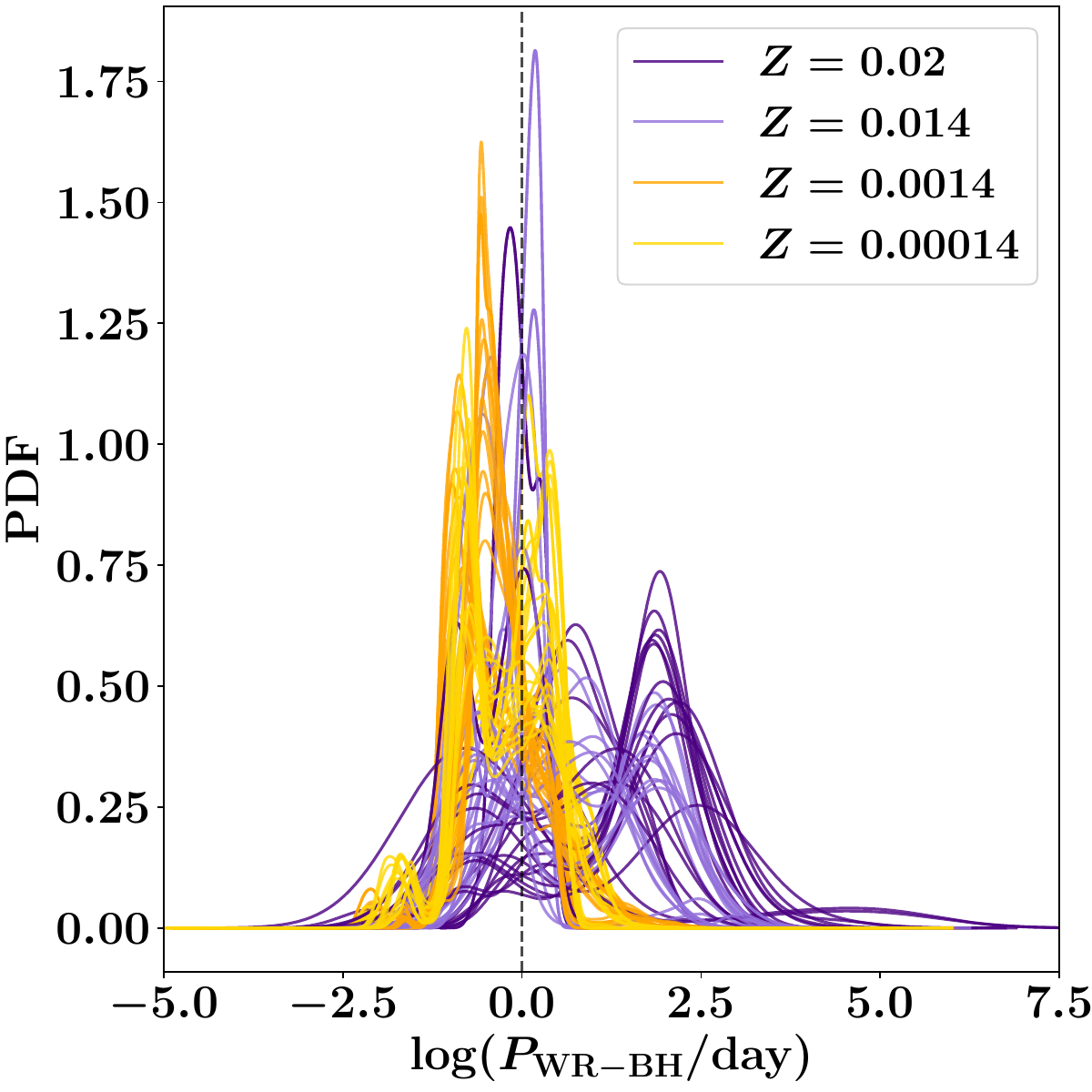}
        \hfill
	\includegraphics[width=0.328\textwidth]{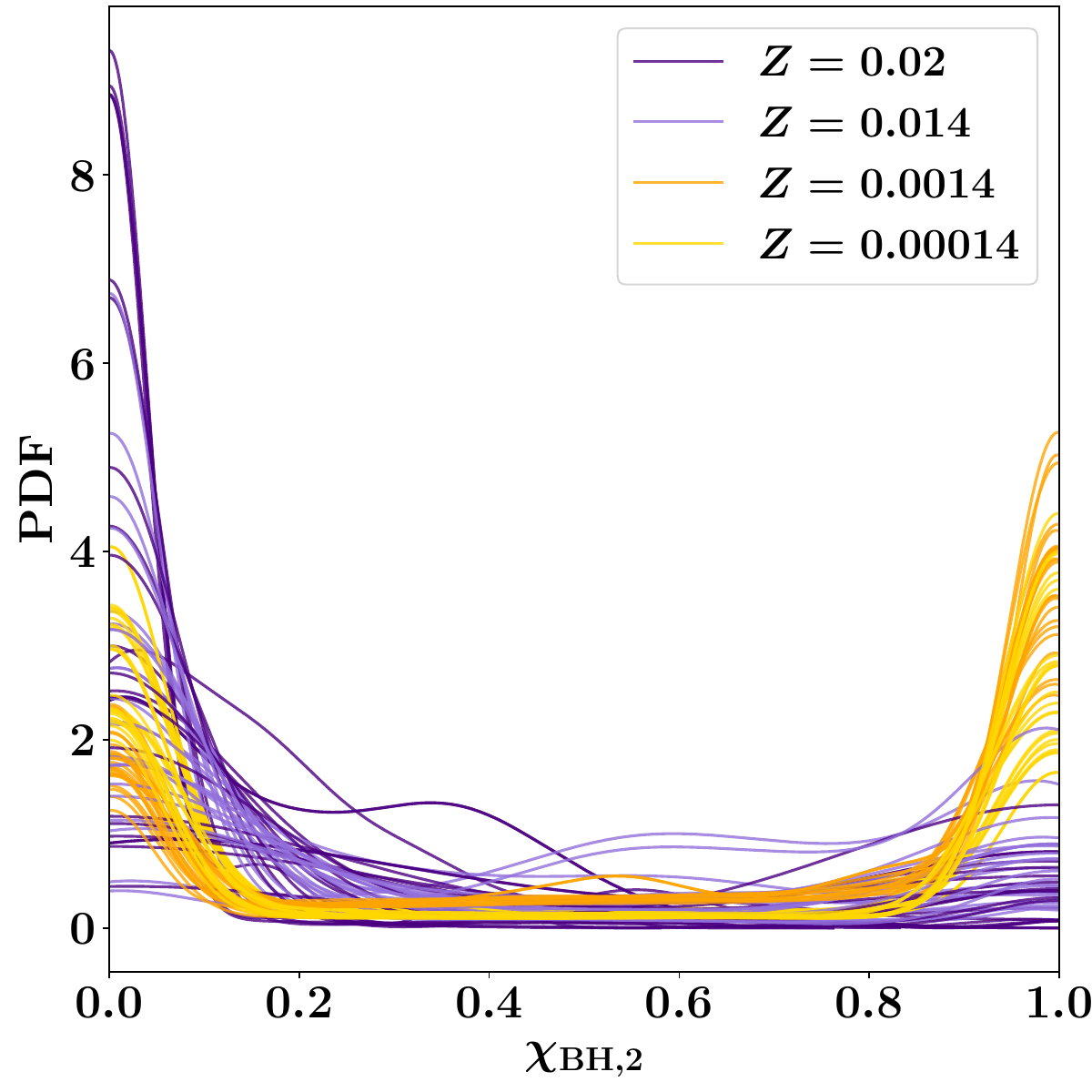}
        \hfill
 	\includegraphics[width=0.328\textwidth]{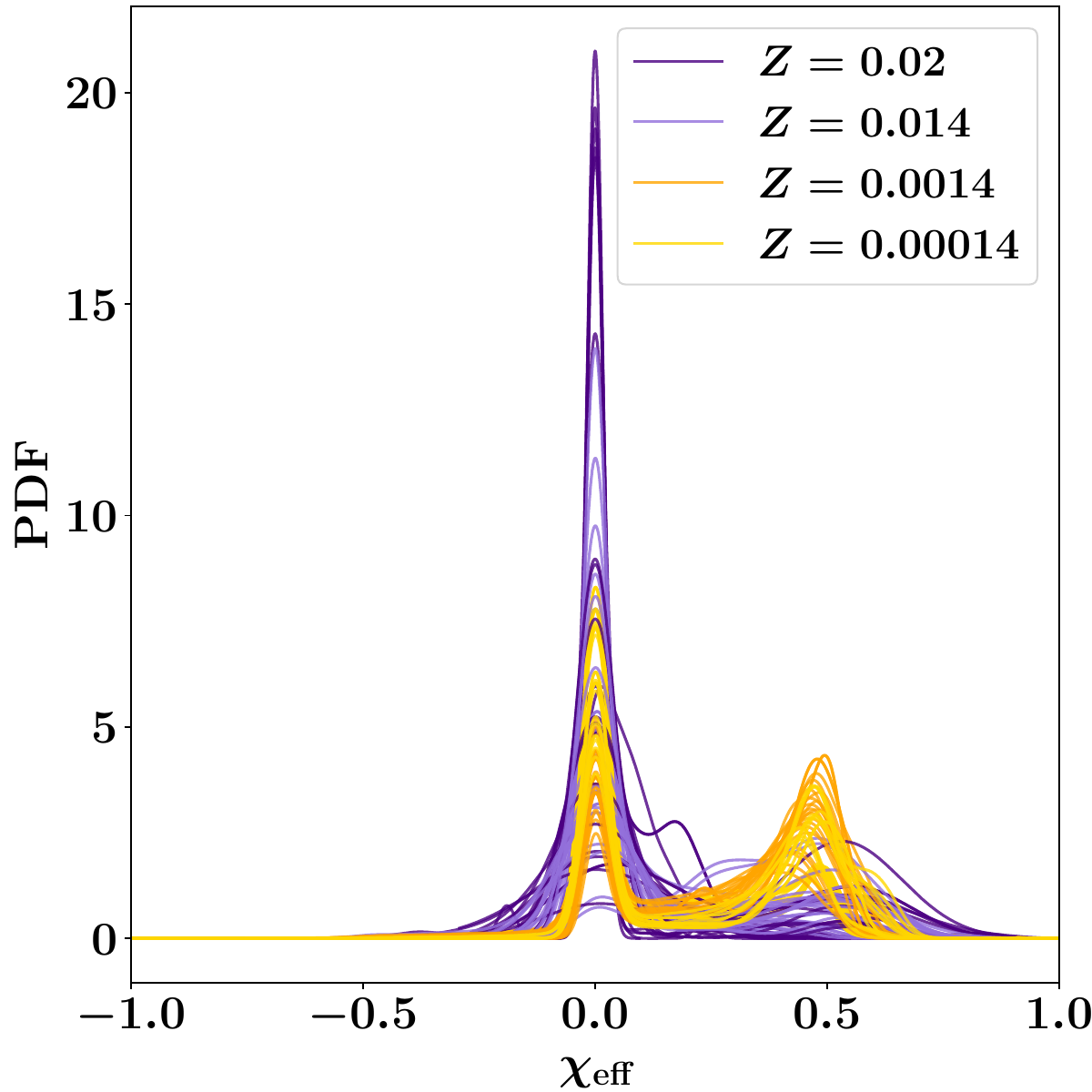}
	\caption{\emph{Left:} Probability distribution function (PDF) of the orbital periods $\rm P_{\rm WR-BH}$ at the end of the WR--BH configuration for progenitors of BBHs. The dashed line at $\rm P_{\rm WR-BH} = 1$ day highlights the systems that are tight enough to induce efficient tidal spin-up ($\rm P_{\rm WR-BH} \leq 1$ day) according to the model by \citet{Bavera2021_BHspinfromWR}. \emph{Center:} Probability distribution function of the dimensionless spin magnitude $\chi_{BH,2}$ of second-born BHs in WR--BHs evolving into BBHs. Spins are calculated with the model of \citet{Bavera2021_BHspinfromWR} that assumes tidal spin-up in close binaries and efficient angular momentum transport. All the 96 simulated sets (Table \ref{tab:parameterspace}) exhibit a bimodal distribution. \emph{Right:} Probability distribution function of the effective spin parameter $\chi_{\rm eff}$ obtained in all 96 sets, color-coded according to the assumed metallicity $Z$. 
	}\label{fig:spins}
\end{figure*}

If we apply this relation to the population of WR--BHs in our simulations, we find a bimodal distribution for the spins of second-born BHs (Figure \ref{fig:spins}): they are either highly spinning ($\chi_2 \approx 1$) or no spinning ($\chi_2 \approx 0$). In the model by  \citet{Bavera2021_BHspinfromWR}, the spin magnitude is determined by tidal spin-up. Thus, it is strongly influenced by the orbital period of the WR--BH system. 

Figure \ref{fig:spins} (left panel) shows that most WR--BHs at solar metallicity ($Z=0.02,0.014$) have orbital periods too large ($P_{\rm WR-BH} > 1$ day) to be spun up by tides according to Equation \ref{eqn:baveraspin}. Only BBHs produced with the G20 natal kick model \citep{SNkicksUnified_Giacobbo2020}, the model that generates the lowest kick magnitudes for BHs (Figure \ref{fig:kickdistributions}), exhibit a significant fraction of binaries with $P_{\rm WR-BH} \lesssim 1$ day. 

At sub-solar metallicity ($Z=0.0014,0.00014$), the lower wind efficiency allows WR production only via mass transfer interactions and binary stripping. Hence, WR--BHs at these metallicities have shorter periods resulting in an efficient spin up of the WR that will produce the second-born BH.

\subsubsection{Comparison with spins from GW observations}
If one of the two BHs is non-spinning ($\chi_1 \approx 0$), the effective spin  parameter $\chi_{\rm eff}$ is extremely sensible to the mass ratio $q=M_{\rm BH,L}/M_{\rm BH,H}$ of the lighter-to-heavier BH ($\rm M_{\rm BH,H} \geq M_{\rm BH,L}$). By construction, this\footnote{Other mass ratio definitions could be larger than 1, for instance the donor-to-accretor mass ratio $q=M_{\rm d}/M_{\rm a}$ used to determine mass transfer stability.} mass ratio is $q<1$ and can be used to calculate $\chi_{\rm eff}$ with an explicit dependence on the possibility that the second-born BH $\rm M_{\rm BH,2}$ is the heaviest or the lightest one:

\begin{equation}\label{eqn:chieff_approx}
    \chi_{\rm eff}  \approx \frac{m_2 ~ \chi_2 ~ \cos \theta_2}{m_1 + m_2} \approx
\begin{cases}
\frac{q}{1+q} ~\chi_2 \cos \theta_2 & \rm M_{\rm BH,2} \leq M_{\rm BH,1}\\
\frac{1}{1+q}~ \chi_2 \cos \theta_2 & \rm M_{\rm BH,2} \geq M_{\rm BH,1}
\end{cases}.
\end{equation}

\begin{figure}[H]
        \centering
        \includegraphics[width=0.5\textwidth]{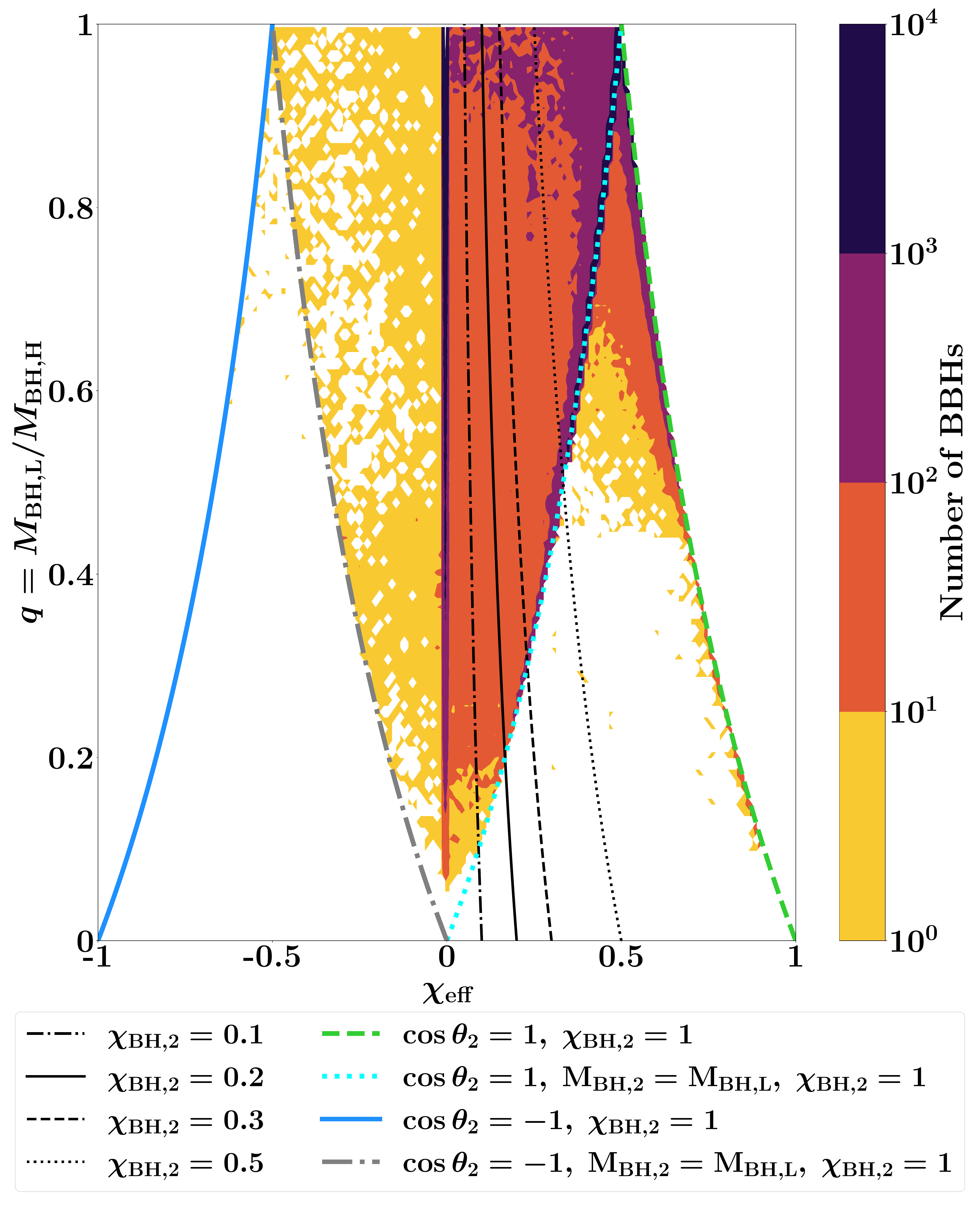}
	\caption{Mass ratio $q=M_{\rm BH,L}/M_{\rm BH,H}$ of the lighter-to-heavier BH} and effective spin parameter $\chi_{\rm eff}$ for all BBHs produced in the 96 sets simulated in this work (Section \ref{sec:methods}). Colored thick lines indicate the analytic relations (Equation \ref{eqn:chieff_approx}) that delimit the allowed parameter space for maximally spinning ($\chi_2 = 1$) second-born BH, for spins aligned ($\cos \theta_2 = 1$) or anti-aligned ($\cos \theta_2 = -1$) with the binary orbital angular momentum. Black thin lines show the analytic relation expected for low spinning ($\chi_2 \sim 0.1-0.5$) and aligned ($\cos \theta_2 = 1$) second-born BH. The analytic relations assume that the second-born BH is the heaviest one ($\rm M_{\rm BH,2} = M_{\rm BH,H}$), unless different specifications. \label{fig:spins_qchieff}
\end{figure}

GW observations suggest that most BHs have relatively low spins ($\chi \approx 0.2$) and that it might exist a $q-\chi_{\rm eff}$ correlation that favors higher $\chi_{\rm eff}$ values for more asymmetric binaries \citep{Callister2021_qVSxeff,GWTC-3_interpretation}. In the case of efficient angular momentum transport and Eddington-limited accretion, the first-born BH could be non-spinning and $\chi_{\rm eff}$ would be determined mainly by the properties of the second-born BH. Assuming also spin aligned to the orbital momentum ($\cos \theta_2 = 1$), the analytical approximation of $\chi_{\rm eff}$ (Equation \ref{eqn:chieff_approx}) indicates an intrinsic $q-\chi_{\rm eff}$ (anti-) correlation if the second-born BH the (most) least massive one. The magnitude of the second-born BH spin influences the shape of the (anti-) correlation in the $q-\chi_{\rm eff}$, as we show in Figure \ref{fig:spins_qchieff}.

In the case of maximally spinning second-born BH ($\chi_{\rm 2} =1$), the approximated formula for $\chi_{\rm eff}$ divides the $q-\chi_{\rm eff}$ plane in three regions: an oval-shaped central region, an outer wing, and a more external one. The regions are symmetric with respect to the $\chi_{\rm eff} = 0$ center, indicating (anti-) aligned spins for (negative) positive $\chi_{\rm eff}$ values. In Figure \ref{fig:spins_qchieff} we superimpose also the properties of all the BBHs simulated in this work (96 sets, see Table \ref{tab:parameterspace}) and spun up via Eq.~\ref{eqn:baveraspin}. We find no BBH in the more external region, thus, no systems with high $\chi_{\rm eff} \approx 1$ and similar masses ($q \approx 1$). Most of our BHs are non-spinning ($\chi_{\rm eff} = 0$) or have spin aligned with the orbital momentum ($\cos \theta_2 \approx 1$)\footnote{The spin orientation $\cos(\theta_2)$ was calculated following Equation 1.16 of \citet{Mapelli2021_reviewBBHformulacostheta}, assuming that only the natal kicks can change the spin orientation.}. BBHs with second-born BH that is lighter than the first-born one ($M_{\rm BH,2} = M_{\rm BH,L}$) are confined in the central oval-shaped region, with $|q| < 0.5$. BBHs with heavier second-born BHs ($M_{\rm BH,2} = M_{\rm BH,H}$) distribute both in the central oval-shaped region and in the outer wings. In this scenario, the wings centered at $|q| = 0.5$ identify a region that can be populated only by BBHs evolved through a mass-ratio-reversal evolution. The distribution of our BBHs in the $q-\chi_{\rm eff}$ plane is in agreement with other works in the literature that studied the mass-ratio-reversal scenario and considered tidal spin up
\citep[e.g.,][]{Olejak2021_BBHspin, Broekgaarden2022_qreversalBBHspin, Olejak2024_BBHspinsMTstable,BanerjeeOlejak2024_BBHspinSMT}.

\section{Summary}\label{sec:conclusions}
We investigated the role of WR--COs as the possible progenitors of BCOs. We used the population synthesis code \sevn to account for uncertainties in theoretical models, evolving the same initial binary population through 96 combinations of metallicity, CE efficiency, CCSN, and natal kick models.

We found that most ($\gtrsim 83\%$) of the simulated BCOs evolved as WR--CO systems in the considered parameter space. More than $\gtrsim 99\%$ ($\approx 83-95 \%$) of BCOs at $Z\geq 0.0014$ ($Z=0.00014$) have a WR--CO progenitor. The fraction of BCOs that derive from WR--CO systems is  mostly affected by mass transfer (when it stops before complete  envelope stripping we do not form a WR at low $Z$), metallicity (which influences stellar winds, thus the role of self and binary stripping in WR production), and the combined effect of natal kicks and CCSN models.

Only $\approx 1-5\%$ of the simulated massive binaries passes through the WR--CO phase. Despite their key role as BCO progenitors, only a small fraction of WR--COs ($\approx 5-30 \%$, depending on the metallicity) becomes a BCO. When formed, WR--COs constitute the last evolutionary configuration prior to BCO formation. These progenitors experience all the RLOs and CE events before the WR--CO phase. Only WR--NSs require at least one additional CE event in order to produce BCOs.

We characterized the possible fate of Cyg X-3, the only Galactic WR--CO candidate, finding that it will likely evolve into a BCO if it is a WR--BH system ($\approx 70-100\%$). If Cyg X-3 is a WR--NS binary, the BCO fate is disfavoured (< 40-60 \%), in agreement with \citet{Belczynski2013_CygX-3fate}. Cyg X-3 possible formation channels are representative of WR--COs at $Z=0.02,0.014$, suggesting that the observation and characterization of similar WR--CO systems could help us to constrain the models to interpret the BCO formation history.

\begin{acknowledgements}
    EK and MM acknowledge support from the PRIN grant METE under contract No. 2020KB33TP. MM, GC, GI, and MDA acknowledge financial support from the European Research Council for the ERC Consolidator grant DEMOBLACK, under contract no. 770017. EK and MM also acknowledge financial support from the German Excellence Strategy via the Heidelberg Cluster of Excellence (EXC 2181 - 390900948) STRUCTURES. GI acknowledges financial support under the National Recovery and Resilience Plan (NRRP), Mission 4, Component 2, Investment 1.4, - Call for tender No. 3138 of 18/12/2021 of Italian Ministry of University and Research funded by the European Union – NextGenerationEU. GC acknowledges financial support from the Agence Nationale de la Recherche grant POPSYCLE number ANR-19-CE31-0022. Numerical
    calculations have been made possible through a CINECA-INFN (TEONGRAV) agreement, providing access to resources on Leonardo at CINECA. The authors acknowledge support by the state of Baden-W\"urttemberg through bwHPC and the German Research Foundation (DFG) through grants INST 35/1597-1 FUGG and INST 35/1503-1 FUGG.
\end{acknowledgements}

\bibliography{biblio}

\begin{thebibliography}{117}
\expandafter\ifx\csname natexlab\endcsname\relax\def\natexlab#1{#1}\fi

\bibitem[{{Abac} {et~al.}(2024){Abac}, {Abbott}, {Abouelfettouh}, {Acernese},
  {Ackley}, {Adhicary}, {Adhikari}, {Adhikari}, {Adkins}, {Agarwal}, {Agathos},
  {Abchouyeh}, {Aguiar}, {Aguilar}, {Aiello}, {Ain}, {Ajith}, {Ak{\c{c}}ay},
  {Akutsu}, {Albanesi}, {Alfaidi}, {Al-Jodah}, {All{\'e}n{\'e}}, {Allocca},
  {Al-Shammari}, {Altin}, {Alvarez-Lopez}, {Amato}, {Amez-Droz}, {Amorosi},
  {Amra}, {Ananyeva}, {Anderson}, {Anderson}, {Andia}, {Ando}, {Andrade},
  {Andres}, {Andr{\'e}s-Carcasona}, {Andri{\'c}}, {Anglin}, {Ansoldi},
  {Antelis}, {Antier}, {Aoumi}, {Appavuravther}, {Appert}, {Apple}, {Arai},
  {Araya}, {Araya}, {Areeda}, {Argianas}, {Aritomi}, {Armato}, {Arnaud},
  {Arogeti}, {Aronson}, {Arun}, {Ashton}, {Aso}, {Assiduo}, {de Souza Melo},
  {Aston}, {Astone}, {Attadio}, {Aubin}, {Aultoneal}, {Avallone}, {Azrad},
  {Babak}, {Badaracco}, {Badger}, {Bae}, {Bagnasco}, {Bagui}, {Baier},
  {Baiotti}, {Bajpai}, {Baka}, {Ball}, {Ballardin}, {Ballmer}, {Banagiri},
  {Banerjee}, {Bankar}, {Baral}, {Barayoga}, {Barish}, {Barker}, {Barneo},
  {Barone}, {Barr}, {Barsotti}, {Barsuglia}, {Barta}, {Bartoletti}, {Barton},
  {Bartos}, \& et~al.}]{GW230529_GWmassgap}
{Abac}, A.~G., {Abbott}, R., {Abouelfettouh}, I., {et~al.} 2024, \apjl, 970,
  L34

\bibitem[{Abbott {et~al.}(2020)Abbott, Abbott, Abbott, Abraham, Acernese,
  Ackley, Adams, Adhikari, Adya, Affeldt, {et~al.}}]{GW190425}
Abbott, B., Abbott, R., Abbott, T., {et~al.} 2020, The Astrophysical Journal,
  892, L3

\bibitem[{Abbott {et~al.}(2019)Abbott, Abbott, Abbott, Acernese, Ackley, Adams,
  Adams, Addesso, Adhikari, Adya, Affeldt, Agarwal, Agathos, Agatsuma,
  Aggarwal, Aguiar, Aiello, Ain, Ajith, Allen, Allen, Allocca, Aloy, Altin,
  Amato, Ananyeva, Anderson, Anderson, Angelova, Antier, Appert, Arai, Araya,
  Areeda, Ar\`ene, Arnaud, Arun, Ascenzi, Ashton, Ast, Aston, Astone, Atallah,
  Aubin, Aufmuth, Aulbert, AultONeal, Austin, Avila-Alvarez, Babak, Bacon,
  Badaracco, Bader, Bae, Baker, Baldaccini, Ballardin, Ballmer, Banagiri,
  Barayoga, Barclay, Barish, Barker, Barkett, Barnum, Barone, Barr, Barsotti,
  Barsuglia, Barta, Bartlett, Bartos, Bassiri, Basti, Batch, Bawaj, Bayley,
  Bazzan, B\'ecsy, Beer, Bejger, Belahcene, Bell, Beniwal, Bensch, Berger,
  Bergmann, Bernuzzi, Bero, Berry, Bersanetti, Bertolini, Betzwieser, Bhandare,
  Bilenko, Bilgili, Billingsley, Billman, Birch, Birney, Birnholtz, Biscans,
  Biscoveanu, Bisht, Bitossi, Bizouard, Blackburn, Blackman, Blair, Blair,
  Blair, Bloemen, Bock, Bode, Boer, Boetzel, Bogaert, Bohe, Bondu, Bonilla,
  Bonnand, Booker, Boom, Booth, Bork, Boschi, Bose, Bossie, Bossilkov, Bosveld,
  Bouffanais, Bozzi, Bradaschia, Brady, Bramley, Branchesi, Brau, Briant,
  Brighenti, Brillet, Brinkmann, Brisson, Brockill, Brooks, Brown, Brunett,
  Buchanan, Buikema, Bulik, Bulten, Buonanno, Buskulic, Buy, Byer, Cabero,
  Cadonati, Cagnoli, Cahillane, Bustillo, Callister, Calloni, Camp, Canepa,
  Canizares, Cannon, Cao, Cao, Capano, Capocasa, Carbognani, Caride, Carney,
  Carullo, Diaz, Casentini, Caudill, Cavagli\`a, Cavalier, Cavalieri, Cella,
  Cepeda, Cerd\'a-Dur\'an, Cerretani, Cesarini, Chaibi, Chamberlin, Chan, Chao,
  Charlton, Chase, Chassande-Mottin, Chatterjee, Chatziioannou, Cheeseboro,
  Chen, Chen, Chen, Cheng, Chia, Chincarini, Chiummo, Chmiel, Cho, Cho, Chow,
  Christensen, Chu, Chua, Chua, Chung, Chung, Ciani, Ciobanu, Ciolfi, Cipriano,
  Cirelli, Cirone, Clara, Clark, Clearwater, Cleva, Cocchieri, Coccia, Cohadon,
  Cohen, Colla, Collette, Collins, Cominsky, Constancio, Conti, Cooper, Corban,
  Corbitt, Cordero-Carri\'on, Corley, Cornish, Corsi, Cortese, Costa, Cotesta,
  Coughlin, Coughlin, Coulon, Countryman, Couvares, Covas, Cowan, Coward,
  Cowart, Coyne, Coyne, Creighton, Creighton, Cripe, Crowder, Cullen, Cumming,
  Cunningham, Cuoco, Canton, D\'alya, Danilishin, D'Antonio, Danzmann,
  Dasgupta, Costa, Dattilo, Dave, Davier, Davis, Daw, Day, DeBra, Deenadayalan,
  Degallaix, De~Laurentis, Del\'eglise, Del~Pozzo, Demos, Denker, Dent,
  De~Pietri, Derby, Dergachev, De~Rosa, De~Rossi, DeSalvo, de~Varona,
  Dhurandhar, D\'{\i}az, Dietrich, Di~Fiore, Di~Giovanni, Di~Girolamo,
  Di~Lieto, Ding, Di~Pace, Di~Palma, Di~Renzo, Dmitriev, Doctor, Dolique,
  Donovan, Dooley, Doravari, Dorrington, \'Alvarez, Downes, Drago,
  Dreissigacker, Driggers, Du, Dudi, Dupej, Dwyer, Easter, Edo, Edwards,
  Effler, Eggenstein, Ehrens, Eichholz, Eikenberry, Eisenmann, Eisenstein,
  Essick, Estelles, Estevez, Etienne, Etzel, Evans, Evans, Fafone, Fair,
  Fairhurst, Fan, Farinon, Farr, Farr, Fauchon-Jones, Favata, Fays, Fee,
  Fehrmann, Feicht, Fejer, Feng, Fernandez-Galiana, Ferrante, Ferreira,
  Ferrini, Fidecaro, Fiori, Fiorucci, Fishbach, Fisher, Fishner, Fitz-Axen,
  Flaminio, Fletcher, Fong, Font, Forsyth, Forsyth, Fournier, Frasca, Frasconi,
  Frei, Freise, Frey, Frey, Fritschel, Frolov, Fulda, Fyffe, Gabbard, Gadre,
  Gaebel, Gair, Gammaitoni, Ganija, Gaonkar, Garcia, Garc\'{\i}a-Quir\'os,
  Garufi, Gateley, Gaudio, Gaur, Gayathri, Gemme, Genin, Gennai, George,
  George, Gergely, Germain, Ghonge, Ghosh, Ghosh, Ghosh, Giacomazzo, Giaime,
  Giardina, Giazotto, Gill, Giordano, Glover, Goetz, Goetz, Goncharov,
  Gonz\'alez, Castro, Gopakumar, Gorodetsky, Gossan, Gosselin, Gouaty, Grado,
  Graef, Granata, Grant, Gras, Gray, Greco, Green, Green, Gretarsson, Groot,
  Grote, Grunewald, Gruning, Guidi, Gulati, Guo, Gupta, Gupta, Gushwa,
  Gustafson, Gustafson, Halim, Hall, Hall, Hamilton, Hamilton, Hammond, Haney,
  Hanke, Hanks, Hanna, Hannam, Hannuksela, Hanson, Hardwick, Harms, Harry,
  Harry, Hart, Haster, Haughian, Healy, Heidmann, Heintze, Heitmann, Hello,
  Hemming, Hendry, Heng, Hennig, Heptonstall, Hernandez, Heurs, Hild, Hinderer,
  Hoak, Hochheim, Hofman, Holland, Holt, Holz, Hopkins, Horst, Hough, Houston,
  Howell, Hreibi, Huerta, Huet, Hughey, Hulko, Husa, Huttner, Huynh-Dinh, Iess,
  Indik, Ingram, Inta, Intini, Isa, Isac, Isi, Iyer, Izumi, Jacqmin, Jani,
  Jaranowski, Johnson, Johnson, Jones, Jones, Jonker, Ju, Junker, Kalaghatgi,
  Kalogera, Kamai, Kandhasamy, Kang, Kanner, Kapadia, Karki, Karvinen,
  Kasprzack, Kastaun, Katolik, Katsanevas, Katsavounidis, Katzman, Kaufer,
  Kawabe, Keerthana, K\'ef\'elian, Keitel, Kemball, Kennedy, Key, Khalili,
  Khamesra, Khan, Khan, Khan, Khan, Khazanov, Kijbunchoo, Kim, Kim, Kim, Kim,
  Kim, Kim, King, King, Kinley-Hanlon, Kirchhoff, Kissel, Kleybolte, Klimenko,
  Knowles, Koch, Koehlenbeck, Koley, Kondrashov, Kontos, Korobko, Korth,
  Kowalska, Kozak, Kr\"amer, Kringel, Krishnan, Kr\'olak, Kuehn, Kumar, Kumar,
  Kumar, Kuo, Kutynia, Kwang, Lackey, Lai, Landry, Landry, Lang, Lange, Lantz,
  Lanza, Lartaux-Vollard, Lasky, Laxen, Lazzarini, Lazzaro, Leaci, Leavey, Lee,
  Lee, Lee, Lee, Lee, Lehmann, Lenon, Leonardi, Leroy, Letendre, Levin, Li, Li,
  Li, Linker, Littenberg, Liu, Liu, Lo, Lockerbie, London, Longo, Lorenzini,
  Loriette, Lormand, Losurdo, Lough, Lousto, Lovelace, L\"uck, Lumaca,
  Lundgren, Lynch, Ma, Macas, Macfoy, Machenschalk, MacInnis, Macleod,
  Hernandez, Maga\~na Sandoval, Zertuche, Magee, Majorana, Maksimovic, Man,
  Mandic, Mangano, Mansell, Manske, Mantovani, Marchesoni, Marion, M\'arka,
  M\'arka, Markakis, Markosyan, Markowitz, Maros, Marquina, Martelli,
  Martellini, Martin, Martin, Martynov, Mason, Massera, Masserot, Massinger,
  Masso-Reid, Mastrogiovanni, Matas, Matichard, Matone, Mavalvala, Mazumder,
  McCann, McCarthy, McClelland, McCormick, McCuller, McGuire, McIver, McManus,
  McRae, McWilliams, Meacher, Meadors, Mehmet, Meidam, Mejuto-Villa, Melatos,
  Mendell, Mendoza-Gandara, Mercer, Mereni, Merilh, Merzougui, Meshkov,
  Messenger, Messick, Metzdorff, Meyers, Miao, Michel, Middleton, Mikhailov,
  Milano, Miller, Miller, Miller, Miller, Millhouse, Mills, Milovich-Goff,
  Minazzoli, Minenkov, Ming, Mishra, Mitra, Mitrofanov, Mitselmakher,
  Mittleman, Moffa, Mogushi, Mohan, Mohapatra, Montani, Moore, Moraru, Moreno,
  Morisaki, Mours, Mow-Lowry, Mueller, Muir, Mukherjee, Mukherjee, Mukherjee,
  Mukund, Mullavey, Munch, Mu\~niz, Muratore, Murray, Nagar, Napier,
  Nardecchia, Naticchioni, Nayak, Neilson, Nelemans, Nelson, Nery, Neunzert,
  Nevin, Newport, Ng, Ng, Nguyen, Nguyen, Nichols, Nielsen, Nissanke, Nitz,
  Nocera, Nolting, North, Nuttall, Obergaulinger, Oberling, O'Brien, O'Dea,
  Ogin, Oh, Oh, Ohme, Ohta, Okada, Oliver, Oppermann, Oram, O'Reilly, Ormiston,
  Ortega, O'Shaughnessy, Ossokine, Ottaway, Overmier, Owen, Pace, Pagano, Page,
  Page, Pai, Pai, Palamos, Palashov, Palomba, Pal-Singh, Pan, Pan, Pang, Pang,
  Pankow, Pannarale, Pant, Paoletti, Paoli, Papa, Parida, Parker, Pascucci,
  Pasqualetti, Passaquieti, Passuello, Patil, Patricelli, Pearlstone, Pedersen,
  Pedraza, Pedurand, Pekowsky, Pele, Penn, Perez, Perreca, Perri, Pfeiffer,
  Phelps, Phukon, Piccinni, Pichot, Piergiovanni, Pierro, Pillant, Pinard,
  Pinto, Pirello, Pitkin, Poggiani, Popolizio, Porter, Possenti, Post, Powell,
  Prasad, Pratt, Pratten, Predoi, Prestegard, Principe, Privitera, Prodi,
  Prokhorov, Puncken, Punturo, Puppo, P\"urrer, Qi, Quetschke, Quintero,
  Quitzow-James, Raab, Rabeling, Radkins, Raffai, Raja, Rajan, Rajbhandari,
  Rakhmanov, Ramirez, Ramos-Buades, Rana, Rapagnani, Raymond, Razzano, Read,
  Regimbau, Rei, Reid, Reitze, Ren, Ricci, Ricker, Riemenschneider, Riles,
  Rizzo, Robertson, Robie, Robinet, Robson, Rocchi, Rolland, Rollins, Roma,
  Romano, Romel, Romie, Rosi\ifmmode~\acute{n}\else \'{n}\fi{}ska, Ross, Rowan,
  R\"udiger, Ruggi, Rutins, Ryan, Sachdev, Sadecki, Sakellariadou, Salconi,
  Saleem, Salemi, Samajdar, Sammut, Sampson, Sanchez, Sanchez, Sanchis-Gual,
  Sandberg, Sanders, Sarin, Sassolas, Sathyaprakash, Saulson, Sauter, Savage,
  Sawadsky, Schale, Scheel, Scheuer, Schmidt, Schnabel, Schofield, Sch\"onbeck,
  Schreiber, Schuette, Schulte, Schutz, Schwalbe, Scott, Scott, Seidel,
  Sellers, Sengupta, Sentenac, Sequino, Sergeev, Setyawati, Shaddock, Shaffer,
  Shah, Shahriar, Shaner, Shao, Shapiro, Shawhan, Shen, Shoemaker, Shoemaker,
  Siellez, Siemens, Sieniawska, Sigg, Silva, Singer, Singh, Singhal, Sintes,
  Slagmolen, Slaven-Blair, Smith, Smith, Smith, Somala, Son, Sorazu,
  Sorrentino, Souradeep, Spencer, Srivastava, Staats, Steinke, Steinlechner,
  Steinlechner, Steinmeyer, Steltner, Stevenson, Stocks, Stone, Stops, Strain,
  Stratta, Strigin, Strunk, Sturani, Stuver, Summerscales, Sun, Sunil, Suresh,
  Sutton, Swinkels, Szczepa\ifmmode~\acute{n}\else \'{n}\fi{}czyk, Tacca, Tait,
  Talbot, Talukder, Tanner, T\'apai, Taracchini, Tasson, Taylor, Taylor,
  Tewari, Theeg, Thies, Thomas, Thomas, Thomas, Thorne, Thrane, Tiwari, Tiwari,
  Tokmakov, Toland, Tonelli, Tornasi, Torres-Forn\'e, Torrie, T\"oyr\"a,
  Travasso, Traylor, Trinastic, Tringali, Trozzo, Tsang, Tse, Tso, Tsuna,
  Tsukada, Tuyenbayev, Ueno, Ugolini, Urban, Usman, Vahlbruch, Vajente, Valdes,
  van Bakel, van Beuzekom, van~den Brand, Van Den~Broeck, Vander-Hyde, van~der
  Schaaf, van Heijningen, van Veggel, Vardaro, Varma, Vass, Vas\'uth, Vecchio,
  Vedovato, Veitch, Veitch, Venkateswara, Venugopalan, Verkindt, Vetrano,
  Vicer\'e, Viets, Vinciguerra, Vine, Vinet, Vitale, Vo, Vocca, Vorvick,
  Vyatchanin, Wade, Wade, Wade, Walet, Walker, Wallace, Walsh, Wang, Wang,
  Wang, Wang, Wang, Ward, Warner, Was, Watchi, Weaver, Wei, Weinert, Weinstein,
  Weiss, Wellmann, Wen, Wessel, We\ss{}els, Westerweck, Wette, Whelan, Whiting,
  Whittle, Wilken, Williams, Williams, Williamson, Willis, Willke, Wimmer,
  Winkler, Wipf, Wittel, Woan, Woehler, Wofford, Wong, Worden, Wright, Wu,
  Wysocki, Xiao, Yam, Yamamoto, Yancey, Yang, Yap, Yazback, Yu, Yu, Yvert,
  Zadro\ifmmode~\dot{z}\else \.{z}\fi{}ny, Zanolin, Zelenova, Zendri, Zevin,
  Zhang, Zhang, Zhang, Zhang, Zhang, Zhao, Zhou, Zhou, Zhu, Zhu, Zimmerman,
  Zlochower, Zucker, \& Zweizig}]{GW170817}
Abbott, B.~P., Abbott, R., Abbott, T.~D., {et~al.} 2019, Phys. Rev. X, 9,
  011001

\bibitem[{Abbott {et~al.}(2023{\natexlab{a}})Abbott, Abbott, Acernese, Ackley,
  Adams, Adhikari, Adhikari, Adya, Affeldt, Agarwal, {et~al.}}]{GWTC-3}
Abbott, R., Abbott, T., Acernese, F., {et~al.} 2023{\natexlab{a}}, Physical
  Review X, 13, 041039

\bibitem[{Abbott {et~al.}(2023{\natexlab{b}})Abbott, Abbott, Acernese, Ackley,
  Adams, Adhikari, Adhikari, Adya, Affeldt, Agarwal,
  {et~al.}}]{GWTC-3_interpretation}
Abbott, R., Abbott, T., Acernese, F., {et~al.} 2023{\natexlab{b}}, Physical
  Review X, 13, 011048

\bibitem[{Abbott {et~al.}(2021)Abbott, Abbott, Abraham, Acernese, Ackley,
  Adams, Adams, Adhikari, Adya, Affeldt, {et~al.}}]{GW200105_GW200115_BHNS}
Abbott, R., Abbott, T.~D., Abraham, S., {et~al.} 2021, The Astrophysical
  journal letters, 915, L5

\bibitem[{{Agrawal} {et~al.}(2022){Agrawal}, {Sz{\'e}csi}, {Stevenson},
  {Eldridge}, \& {Hurley}}]{Poojan2022_stellarevodifferences}
{Agrawal}, P., {Sz{\'e}csi}, D., {Stevenson}, S., {Eldridge}, J.~J., \&
  {Hurley}, J. 2022, \mnras, 512, 5717

\bibitem[{{Atri} {et~al.}(2019){Atri}, {Miller-Jones}, {Bahramian}, {Plotkin},
  {Jonker}, {Nelemans}, {Maccarone}, {Sivakoff}, {Deller}, {Chaty}, {Torres},
  {Horiuchi}, {McCallum}, {Natusch}, {Phillips}, {Stevens}, \&
  {Weston}}]{Atri2019_kicks}
{Atri}, P., {Miller-Jones}, J. C.~A., {Bahramian}, A., {et~al.} 2019, \mnras,
  489, 3116

\bibitem[{{Banerjee} \& {Olejak}(2024)}]{BanerjeeOlejak2024_BBHspinSMT}
{Banerjee}, S. \& {Olejak}, A. 2024, Submitted to \aap, arXiv:2411.15112

\bibitem[{Bartzakos {et~al.}(2001)Bartzakos, Moffat, \&
  Niemela}]{Bartzakos2001_WRcatalogueMC}
Bartzakos, P., Moffat, A., \& Niemela, V. 2001, Monthly Notices of the Royal
  Astronomical Society, 324, 18

\bibitem[{Bavera {et~al.}(2021)Bavera, Fragos, Zevin, Berry, Marchant, Andrews,
  Coughlin, Dotter, Kovlakas, Misra, {et~al.}}]{Bavera2021masstransfer}
Bavera, S.~S., Fragos, T., Zevin, M., {et~al.} 2021, Astronomy \& Astrophysics,
  647, A153

\bibitem[{{Bavera} {et~al.}(2021{\natexlab{a}}){Bavera}, {Fragos}, {Zevin},
  {Berry}, {Marchant}, {Andrews}, {Coughlin}, {Dotter}, {Kovlakas}, {Misra},
  {Serra-Perez}, {Qin}, {Rocha}, {Rom{\'a}n-Garza}, {Tran}, \&
  {Zapartas}}]{Bavera2021_BHspinfromWRtheory}
{Bavera}, S.~S., {Fragos}, T., {Zevin}, M., {et~al.} 2021{\natexlab{a}}, \aap,
  647, A153

\bibitem[{{Bavera} {et~al.}(2021{\natexlab{b}}){Bavera}, {Zevin}, \&
  {Fragos}}]{Bavera2021_BHspinfromWR}
{Bavera}, S.~S., {Zevin}, M., \& {Fragos}, T. 2021{\natexlab{b}}, Research
  Notes of the American Astronomical Society, 5, 127

\bibitem[{{Belczynski} {et~al.}(2012){Belczynski}, {Bulik}, \&
  {Fryer}}]{Belczynski2012_HMXBHfate}
{Belczynski}, K., {Bulik}, T., \& {Fryer}, C.~L. 2012, Submitted to \apj,
  arXiv:1208.2422

\bibitem[{Belczynski {et~al.}(2013)Belczynski, Bulik, Mandel, Sathyaprakash,
  Zdziarski, \& Mikolajewska}]{Belczynski2013_CygX-3fate}
Belczynski, K., Bulik, T., Mandel, I., {et~al.} 2013, The Astrophysical
  Journal, 764, 96

\bibitem[{{Belczynski} {et~al.}(2008){Belczynski}, {Kalogera}, {Rasio}, {Taam},
  {Zezas}, {Bulik}, {Maccarone}, \& {Ivanova}}]{Belczynski2008_STARTRACK}
{Belczynski}, K., {Kalogera}, V., {Rasio}, F.~A., {et~al.} 2008, \apjs, 174,
  223

\bibitem[{{Belczynski} {et~al.}(2020){Belczynski}, {Klencki}, {Fields},
  {Olejak}, {Berti}, {Meynet}, {Fryer}, {Holz}, {O'Shaughnessy}, {Brown},
  {Bulik}, {Leung}, {Nomoto}, {Madau}, {Hirschi}, {Kaiser}, {Jones}, {Mondal},
  {Chruslinska}, {Drozda}, {Gerosa}, {Doctor}, {Giersz}, {Ekstrom}, {Georgy},
  {Askar}, {Baibhav}, {Wysocki}, {Natan}, {Farr}, {Wiktorowicz}, {Coleman
  Miller}, {Farr}, \& {Lasota}}]{Belczynski2020_popsynthchannel}
{Belczynski}, K., {Klencki}, J., {Fields}, C.~E., {et~al.} 2020, \aap, 636,
  A104

\bibitem[{{Bethe} \& {Brown}(1998)}]{bethe1998}
{Bethe}, H.~A. \& {Brown}, G.~E. 1998, \apj, 506, 780

\bibitem[{Binder {et~al.}(2021)Binder, Sy, Eracleous, Christodoulou,
  Bhattacharya, Cappallo, Laycock, Plucinsky, \&
  Williams}]{NGC300X-1_Binder2021_BHpreciso}
Binder, B.~A., Sy, J.~M., Eracleous, M., {et~al.} 2021, The Astrophysical
  Journal, 910, 74

\bibitem[{{Boesky} {et~al.}(2024){Boesky}, {Broekgaarden}, \&
  {Berger}}]{Boesky2024}
{Boesky}, A.~P., {Broekgaarden}, F.~S., \& {Berger}, E. 2024, \apj, 976, 23

\bibitem[{{B{\"o}hm-Vitense}(1958)}]{Bohm-Vitense1958}
{B{\"o}hm-Vitense}, E. 1958, \zap, 46, 108

\bibitem[{Bressan {et~al.}(1981)Bressan, Chiosi, \&
  Bertelli}]{Bressan1981_ballisticapproach}
Bressan, A., Chiosi, C., \& Bertelli, G. 1981, Astronomy and Astrophysics, vol.
  102, no. 1, Sept. 1981, p. 25-30. Consiglio Nazionale delle Ricerche, 102, 25

\bibitem[{{Bressan} {et~al.}(2012){Bressan}, {Marigo}, {Girardi}, {Salasnich},
  {Dal Cero}, {Rubele}, \& {Nanni}}]{parsec2012}
{Bressan}, A., {Marigo}, P., {Girardi}, L., {et~al.} 2012, \mnras, 427, 127

\bibitem[{{Broekgaarden} {et~al.}(2022{\natexlab{a}}){Broekgaarden}, {Berger},
  {Stevenson}, {Justham}, {Mandel}, {Chru{\'s}li{\'n}ska}, {van Son}, {Wagg},
  {Vigna-G{\'o}mez}, {de Mink}, {Chattopadhyay}, \&
  {Neijssel}}]{Broekgaarden2022_formationchannels}
{Broekgaarden}, F.~S., {Berger}, E., {Stevenson}, S., {et~al.}
  2022{\natexlab{a}}, \mnras, 516, 5737

\bibitem[{{Broekgaarden} {et~al.}(2022{\natexlab{b}}){Broekgaarden},
  {Stevenson}, \& {Thrane}}]{Broekgaarden2022_qreversalBBHspin}
{Broekgaarden}, F.~S., {Stevenson}, S., \& {Thrane}, E. 2022{\natexlab{b}},
  \apj, 938, 45

\bibitem[{{Callister} {et~al.}(2021){Callister}, {Haster}, {Ng}, {Vitale}, \&
  {Farr}}]{Callister2021_qVSxeff}
{Callister}, T.~A., {Haster}, C.-J., {Ng}, K. K.~Y., {Vitale}, S., \& {Farr},
  W.~M. 2021, \apjl, 922, L5

\bibitem[{{Chen} {et~al.}(2015){Chen}, {Bressan}, {Girardi}, {Marigo}, {Kong},
  \& {Lanza}}]{parsec2015_chen}
{Chen}, Y., {Bressan}, A., {Girardi}, L., {et~al.} 2015, \mnras, 452, 1068

\bibitem[{{Chieffi} \&
  {Limongi}(2020)}]{ChieffiLimongi2020_compactnessnonmonotonic}
{Chieffi}, A. \& {Limongi}, M. 2020, \apj, 890, 43

\bibitem[{{Claeys} {et~al.}(2014){Claeys}, {Pols}, {Izzard}, {Vink}, \&
  {Verbunt}}]{Clayes2014_lambdaCE}
{Claeys}, J. S.~W., {Pols}, O.~R., {Izzard}, R.~G., {Vink}, J., \& {Verbunt},
  F. W.~M. 2014, \aap, 563, A83

\bibitem[{{Costa} {et~al.}(2021){Costa}, {Bressan}, {Mapelli}, {Marigo},
  {Iorio}, \& {Spera}}]{MassGapStellarEvo_Costa2021}
{Costa}, G., {Bressan}, A., {Mapelli}, M., {et~al.} 2021, \mnras, 501, 4514

\bibitem[{{Costa} {et~al.}(2019){Costa}, {Girardi}, {Bressan}, {Marigo},
  {Rodrigues}, {Chen}, {Lanza}, \& {Goudfrooij}}]{parsec2019Costa}
{Costa}, G., {Girardi}, L., {Bressan}, A., {et~al.} 2019, \mnras, 485, 4641

\bibitem[{{Costa} {et~al.}(2025){Costa}, {Shepherd}, {Bressan}, {Addari},
  {Chen}, {Fu}, {Volpato}, {Nguyen}, {Girardi}, {Marigo}, {Mazzi},
  {Pastorelli}, {Trabucchi}, {Bossini}, \& {Zaggia}}]{Costa2025_PARSECtracks}
{Costa}, G., {Shepherd}, K.~G., {Bressan}, A., {et~al.} 2025, Accepted for
  publication in \aap, arXiv:2501.12917

\bibitem[{Crowther(2007)}]{Crowther2007_reviewWR}
Crowther, P.~A. 2007, Annu. Rev. Astron. Astrophys., 45, 177

\bibitem[{{Dall'Amico} {et~al.}(2025){Dall'Amico}, {Mapelli}, {Iorio}, {Costa},
  {Charlot}, {Korb}, {Sgalletta}, \& {Lecroq}}]{DallAmico2025}
{Dall'Amico}, M., {Mapelli}, M., {Iorio}, G., {et~al.} 2025, Submitted to \aap,
  arXiv:2501.04778

\bibitem[{{Dominik} {et~al.}(2012){Dominik}, {Belczynski}, {Fryer}, {Holz},
  {Berti}, {Bulik}, {Mandel}, \& {O'Shaughnessy}}]{Dominik2012}
{Dominik}, M., {Belczynski}, K., {Fryer}, C., {et~al.} 2012, \apj, 759, 52

\bibitem[{{Esposito} {et~al.}(2015){Esposito}, {Israel}, {Milisavljevic},
  {Mapelli}, {Zampieri}, {Sidoli}, {Fabbiano}, \& {Rodr{\'\i}guez
  Castillo}}]{observations}
{Esposito}, P., {Israel}, G.~L., {Milisavljevic}, D., {et~al.} 2015, \mnras,
  452, 1112

\bibitem[{{Fern{\'a}ndez} {et~al.}(2018){Fern{\'a}ndez}, {Quataert},
  {Kashiyama}, \& {Coughlin}}]{fernandez2018}
{Fern{\'a}ndez}, R., {Quataert}, E., {Kashiyama}, K., \& {Coughlin}, E.~R.
  2018, \mnras, 476, 2366

\bibitem[{Foellmi {et~al.}(2003{\natexlab{a}})Foellmi, Moffat, \&
  Guerrero}]{Foellmi2003_WRcatalogueSMC}
Foellmi, C., Moffat, A., \& Guerrero, M. 2003{\natexlab{a}}, Monthly Notices of
  the Royal Astronomical Society, 338, 360

\bibitem[{Foellmi {et~al.}(2003{\natexlab{b}})Foellmi, Moffat, \&
  Guerrero}]{Foellmi2003_WRcatalogueLMC}
Foellmi, C., Moffat, A., \& Guerrero, M. 2003{\natexlab{b}}, Monthly Notices of
  the Royal Astronomical Society, 338, 1025

\bibitem[{{Fryer} {et~al.}(2012){Fryer}, {Belczynski}, {Wiktorowicz},
  {Dominik}, {Kalogera}, \& {Holz}}]{Fryer2012}
{Fryer}, C.~L., {Belczynski}, K., {Wiktorowicz}, G., {et~al.} 2012, \apj, 749,
  91

\bibitem[{{Gallegos-Garcia} {et~al.}(2021){Gallegos-Garcia}, {Berry},
  {Marchant}, \& {Kalogera}}]{gallegos2021MESAvspopsynth}
{Gallegos-Garcia}, M., {Berry}, C. P.~L., {Marchant}, P., \& {Kalogera}, V.
  2021, \apj, 922, 110

\bibitem[{{Ge} {et~al.}(2010){Ge}, {Hjellming}, {Webbink}, {Chen}, \&
  {Han}}]{Ge2010}
{Ge}, H., {Hjellming}, M.~S., {Webbink}, R.~F., {Chen}, X., \& {Han}, Z. 2010,
  \apj, 717, 724

\bibitem[{{Ge} {et~al.}(2024){Ge}, {Tout}, {Chen}, {Wang}, {Xiong}, {Zhang},
  {Li}, {Liu}, \& {Han}}]{Ge2024}
{Ge}, H., {Tout}, C.~A., {Chen}, X., {et~al.} 2024, \apj, 975, 254

\bibitem[{{Ge} {et~al.}(2015){Ge}, {Webbink}, {Chen}, \& {Han}}]{Ge2015}
{Ge}, H., {Webbink}, R.~F., {Chen}, X., \& {Han}, Z. 2015, \apj, 812, 40

\bibitem[{{Ge} {et~al.}(2020){Ge}, {Webbink}, {Chen}, \& {Han}}]{Ge2020}
{Ge}, H., {Webbink}, R.~F., {Chen}, X., \& {Han}, Z. 2020, \apj, 899, 132

\bibitem[{{Giacobbo} \& {Mapelli}(2018)}]{giacobbomapelli2018_mobse_fryer}
{Giacobbo}, N. \& {Mapelli}, M. 2018, \mnras, 480, 2011

\bibitem[{{Giacobbo} \& {Mapelli}(2020)}]{SNkicksUnified_Giacobbo2020}
{Giacobbo}, N. \& {Mapelli}, M. 2020, \apj, 891, 141

\bibitem[{{G{\"o}tberg} {et~al.}(2023){G{\"o}tberg}, {Drout}, {Ji}, {Groh},
  {Ludwig}, {Crowther}, {Smith}, {de Koter}, \& {de
  Mink}}]{Gotberg23_newWRobservations_2307.00074}
{G{\"o}tberg}, Y., {Drout}, M.~R., {Ji}, A.~P., {et~al.} 2023, \apj, 959, 125

\bibitem[{{Gr{\"a}fener} \& {Hamann}(2008)}]{GrafenerHamann2008_WRwinds}
{Gr{\"a}fener}, G. \& {Hamann}, W.~R. 2008, \aap, 482, 945

\bibitem[{{Gr{\"a}fener} {et~al.}(2017){Gr{\"a}fener}, {Owocki}, {Grassitelli},
  \& {Langer}}]{Grafener2017_WRwindssonicpoint}
{Gr{\"a}fener}, G., {Owocki}, S.~P., {Grassitelli}, L., \& {Langer}, N. 2017,
  \aap, 608, A34

\bibitem[{{Gr{\"a}fener} \&
  {Vink}(2013)}]{GrafenerVink2013_WRwindsTemperatureRadiusDef}
{Gr{\"a}fener}, G. \& {Vink}, J.~S. 2013, \aap, 560, A6

\bibitem[{{Hainich} {et~al.}(2014){Hainich}, {R{\"u}hling}, {Todt}, {Oskinova},
  {Liermann}, {Gr{\"a}fener}, {Foellmi}, {Schnurr}, \&
  {Hamann}}]{Hainich2014_WRobservationPOWRgrids}
{Hainich}, R., {R{\"u}hling}, U., {Todt}, H., {et~al.} 2014, \aap, 565, A27

\bibitem[{{Hamann} \& {Gr{\"a}fener}(2004)}]{HamannGraefener2004_powrWRwinds}
{Hamann}, W.~R. \& {Gr{\"a}fener}, G. 2004, \aap, 427, 697

\bibitem[{{Han} {et~al.}(1994){Han}, {Podsiadlowski}, \& {Eggleton}}]{Han1994}
{Han}, Z., {Podsiadlowski}, P., \& {Eggleton}, P.~P. 1994, \mnras, 270, 121

\bibitem[{{Han} {et~al.}(1995){Han}, {Podsiadlowski}, \&
  {Eggleton}}]{Han1995_CErecombinationenergy}
{Han}, Z., {Podsiadlowski}, P., \& {Eggleton}, P.~P. 1995, \mnras, 272, 800

\bibitem[{{Herwig} {et~al.}(1997){Herwig}, {Bloecker}, {Schoenberner}, \& {El
  Eid}}]{Herwig1997_overshooting}
{Herwig}, F., {Bloecker}, T., {Schoenberner}, D., \& {El Eid}, M. 1997, \aap,
  324, L81

\bibitem[{{Hjellming} \& {Webbink}(1987)}]{hjellmingwebbink1987_coreRLOF}
{Hjellming}, M.~S. \& {Webbink}, R.~F. 1987, \apj, 318, 794

\bibitem[{{Hobbs} {et~al.}(2005){Hobbs}, {Lorimer}, {Lyne}, \&
  {Kramer}}]{Hobbs2005}
{Hobbs}, G., {Lorimer}, D.~R., {Lyne}, A.~G., \& {Kramer}, M. 2005, \mnras,
  360, 974

\bibitem[{{Hurley} {et~al.}(2002){Hurley}, {Tout}, \& {Pols}}]{Hurley2002}
{Hurley}, J.~R., {Tout}, C.~A., \& {Pols}, O.~R. 2002, \mnras, 329, 897

\bibitem[{{Iorio} {et~al.}(2023){Iorio}, {Mapelli}, {Costa}, {Spera},
  {Escobar}, {Sgalletta}, {Trani}, {Korb}, {Santoliquido}, {Dall'Amico},
  {Gaspari}, \& {Bressan}}]{Iorio2023_SEVN2}
{Iorio}, G., {Mapelli}, M., {Costa}, G., {et~al.} 2023, \mnras, 524, 426

\bibitem[{{Ivanova} {et~al.}(2013){Ivanova}, {Justham}, {Chen}, {De Marco},
  {Fryer}, {Gaburov}, {Ge}, {Glebbeek}, {Han}, {Li}, {Lu}, {Marsh},
  {Podsiadlowski}, {Potter}, {Soker}, {Taam}, {Tauris}, {van den Heuvel}, \&
  {Webbink}}]{Ivanova2013_CE}
{Ivanova}, N., {Justham}, S., {Chen}, X., {et~al.} 2013, \aapr, 21, 59

\bibitem[{Klencki {et~al.}(2021)Klencki, Nelemans, Istrate, \&
  Chruslinska}]{Klencki2021CE}
Klencki, J., Nelemans, G., Istrate, A.~G., \& Chruslinska, M. 2021, Astronomy
  \& Astrophysics, 645, A54

\bibitem[{{Koljonen} \& {Maccarone}(2017)}]{CygX-3_Koljonen2017}
{Koljonen}, K. I.~I. \& {Maccarone}, T.~J. 2017, \mnras, 472, 2181

\bibitem[{Kroupa(2001)}]{Kroupa2001}
Kroupa, P. 2001, Monthly Notices of the Royal Astronomical Society, 322, 231

\bibitem[{Laplace {et~al.}(2020)Laplace, G{\"o}tberg, De~Mink, Justham, \&
  Farmer}]{Laplace2020_WRradius}
Laplace, E., G{\"o}tberg, Y., De~Mink, S., Justham, S., \& Farmer, R. 2020,
  Astronomy \& Astrophysics, 637, A6

\bibitem[{{Laycock} {et~al.}(2015){Laycock}, {Maccarone}, \&
  {Christodoulou}}]{ICX10X-1_Laycock2015_revisited}
{Laycock}, S. G.~T., {Maccarone}, T.~J., \& {Christodoulou}, D.~M. 2015,
  \mnras, 452, L31

\bibitem[{{Lemasle} {et~al.}(2018){Lemasle}, {Hajdu}, {Kovtyukh}, {Inno},
  {Grebel}, {Catelan}, {Bono}, {Fran{\c{c}}ois}, {Kniazev}, {da Silva}, \&
  {Storm}}]{MetallicityGradientMW2018Gaia}
{Lemasle}, B., {Hajdu}, G., {Kovtyukh}, V., {et~al.} 2018, \aap, 618, A160

\bibitem[{{Limongi} \& {Chieffi}(2010)}]{Limongi2010_preSNevo}
{Limongi}, M. \& {Chieffi}, A. 2010, in Journal of Physics Conference Series,
  Vol. 202, Journal of Physics Conference Series, 012002

\bibitem[{{Limongi} \& {Chieffi}(2018)}]{Limongi2018_rotatingCOcompactness}
{Limongi}, M. \& {Chieffi}, A. 2018, \apjs, 237, 13

\bibitem[{{Mandel} \&
  {Broekgaarden}(2022)}]{MandelBroekgaarden2022_modeluncertainties}
{Mandel}, I. \& {Broekgaarden}, F.~S. 2022, Living Reviews in Relativity, 25, 1

\bibitem[{{Mapelli}(2021)}]{Mapelli2021_reviewBBHformulacostheta}
{Mapelli}, M. 2021, in Handbook of Gravitational Wave Astronomy, 16

\bibitem[{{Mapelli} {et~al.}(2020){Mapelli}, {Spera}, {Montanari}, {Limongi},
  {Chieffi}, {Giacobbo}, {Bressan}, \& {Bouffanais}}]{mapelli2020_compactness}
{Mapelli}, M., {Spera}, M., {Montanari}, E., {et~al.} 2020, \apj, 888, 76

\bibitem[{McCollough {et~al.}(2016)McCollough, Corrales, \&
  Dunham}]{CygX-3_McCollough2016_Observation}
McCollough, M., Corrales, L., \& Dunham, M. 2016, The Astrophysical Journal
  Letters, 830, L36

\bibitem[{Moe \& Di~Stefano(2017)}]{MoeDiStefano2017}
Moe, M. \& Di~Stefano, R. 2017, The Astrophysical Journal Supplement Series,
  230, 15

\bibitem[{{Neijssel} {et~al.}(2019){Neijssel}, {Vigna-G{\'o}mez}, {Stevenson},
  {Barrett}, {Gaebel}, {Broekgaarden}, {de Mink}, {Sz{\'e}csi}, {Vinciguerra},
  \& {Mandel}}]{Neijssel2019}
{Neijssel}, C.~J., {Vigna-G{\'o}mez}, A., {Stevenson}, S., {et~al.} 2019,
  \mnras, 490, 3740

\bibitem[{{Nguyen} {et~al.}(2022){Nguyen}, {Costa}, {Girardi}, {Volpato},
  {Bressan}, {Chen}, {Marigo}, {Fu}, \&
  {Goudfrooij}}]{parsec2022_Nguyenrotation}
{Nguyen}, C.~T., {Costa}, G., {Girardi}, L., {et~al.} 2022, \aap, 665, A126

\bibitem[{{Nugis} \& {Lamers}(2000)}]{Nugis2000_WRwinds}
{Nugis}, T. \& {Lamers}, H. J. G. L.~M. 2000, \aap, 360, 227

\bibitem[{{Nugis} \& {Lamers}(2002)}]{NugisLamers2002_WRwinds}
{Nugis}, T. \& {Lamers}, H.~J.~G.~L.~M. 2002, \aap, 389, 162

\bibitem[{{O'Connor} \& {Ott}(2011)}]{Oconnor2011_compactness}
{O'Connor}, E. \& {Ott}, C.~D. 2011, \apj, 730, 70

\bibitem[{{Olejak} \& {Belczynski}(2021)}]{Olejak2021_BBHspin}
{Olejak}, A. \& {Belczynski}, K. 2021, \apjl, 921, L2

\bibitem[{{Olejak} {et~al.}(2024){Olejak}, {Klencki}, {Xu}, {Wang},
  {Belczynski}, \& {Lasota}}]{Olejak2024_BBHspinsMTstable}
{Olejak}, A., {Klencki}, J., {Xu}, X.-T., {et~al.} 2024, \aap, 689, A305

\bibitem[{{{\"O}zel} \&
  {Freire}(2016)}]{OzelFreire2016_NSmassreviewobservations}
{{\"O}zel}, F. \& {Freire}, P. 2016, \araa, 54, 401

\bibitem[{{Paczy{\'n}ski} \& {Zi{\'o}{\l}kowski}(1968)}]{Paczynski1968}
{Paczy{\'n}ski}, B. \& {Zi{\'o}{\l}kowski}, J. 1968, \actaa, 18, 255

\bibitem[{{Patton} \&
  {Sukhbold}(2020)}]{PattonSukhbold2020_compactnessnonmonotonic}
{Patton}, R.~A. \& {Sukhbold}, T. 2020, \mnras, 499, 2803

\bibitem[{Paxton {et~al.}(2015)Paxton, Marchant, Schwab, Bauer, Bildsten,
  Cantiello, Dessart, Farmer, Hu, Langer, {et~al.}}]{MESA2015BinaryMT}
Paxton, B., Marchant, P., Schwab, J., {et~al.} 2015, The Astrophysical Journal
  Supplement Series, 220, 15

\bibitem[{Peters(1964)}]{Peters1964}
Peters, P.~C. 1964, Physical Review, 136, B1224

\bibitem[{{Picco} {et~al.}(2024){Picco}, {Marchant}, {Sana}, \&
  {Nelemans}}]{Picco2024}
{Picco}, A., {Marchant}, P., {Sana}, H., \& {Nelemans}, G. 2024, \aap, 681, A31

\bibitem[{{R{\"o}pke} \& {De Marco}(2023)}]{RoepkeDeMarco2022_CEreview}
{R{\"o}pke}, F.~K. \& {De Marco}, O. 2023, Living Reviews in Computational
  Astrophysics, 9, 2

\bibitem[{{Sabhahit} {et~al.}(2022){Sabhahit}, {Vink}, {Higgins}, \&
  {Sander}}]{WRwindstemperaturedependence_Sabhahit2023}
{Sabhahit}, G.~N., {Vink}, J.~S., {Higgins}, E.~R., \& {Sander}, A. A.~C. 2022,
  \mnras, 514, 3736

\bibitem[{Sana {et~al.}(2012)Sana, De~Mink, de~Koter, Langer, Evans, Gieles,
  Gosset, Izzard, Le~Bouquin, \& Schneider}]{Sana2012}
Sana, H., De~Mink, S., de~Koter, A., {et~al.} 2012, Science, 337, 444

\bibitem[{Sander {et~al.}(2019)Sander, Hamann, Todt, Hainich, Shenar,
  Ramachandran, \& Oskinova}]{Sander2019_WRwinds}
Sander, A. A.~C., Hamann, W.-R., Todt, H., {et~al.} 2019, Astronomy \&
  Astrophysics, 621, A92

\bibitem[{{Sander} {et~al.}(2023){Sander}, {Lefever}, {Poniatowski},
  {Ramachandran}, {Sabhahit}, \&
  {Vink}}]{WRwindstemperaturedependence_Sander2023}
{Sander}, A.~A.~C., {Lefever}, R.~R., {Poniatowski}, L.~G., {et~al.} 2023,
  \aap, 670, A83

\bibitem[{{Santoliquido} {et~al.}(2021){Santoliquido}, {Mapelli}, {Giacobbo},
  {Bouffanais}, \& {Artale}}]{Santoliquido2021_mergerrate}
{Santoliquido}, F., {Mapelli}, M., {Giacobbo}, N., {Bouffanais}, Y., \&
  {Artale}, M.~C. 2021, \mnras, 502, 4877

\bibitem[{{Schneider} {et~al.}(2023){Schneider}, {Podsiadlowski}, \&
  {Laplace}}]{Schneider2023_strippedstars}
{Schneider}, F. R.~N., {Podsiadlowski}, P., \& {Laplace}, E. 2023, \apjl, 950,
  L9

\bibitem[{{Schneider} {et~al.}(2021){Schneider}, {Podsiadlowski}, \&
  {M{\"u}ller}}]{Schneider2021_SNfromstrippedstars}
{Schneider}, F.~R.~N., {Podsiadlowski}, P., \& {M{\"u}ller}, B. 2021, \aap,
  645, A5

\bibitem[{Schnurr {et~al.}(2008)Schnurr, Casoli, Chen{\'e}, Moffat, \&
  St-Louis}]{Schnurr2008_WRcatalogueLMC}
Schnurr, O., Casoli, J., Chen{\'e}, A.-N., Moffat, A., \& St-Louis, N. 2008,
  Monthly Notices of the Royal Astronomical Society: Letters, 389, L38

\bibitem[{{Schwarzschild}(1958)}]{Schwarzschild1958}
{Schwarzschild}, M. 1958, {Structure and evolution of the stars.}

\bibitem[{{Sen} {et~al.}(2021){Sen}, {Xu}, {Langer}, {El Mellah},
  {Sch{\"u}rmann}, \& {Quast}}]{KoushikSen2021_WR-OandBH-Obinaries}
{Sen}, K., {Xu}, X.~T., {Langer}, N., {et~al.} 2021, \aap, 652, A138

\bibitem[{{Shenar} {et~al.}(2020){Shenar}, {Gilkis}, {Vink}, {Sana}, \&
  {Sander}}]{Shenar2020_WRstarsisolated}
{Shenar}, T., {Gilkis}, A., {Vink}, J.~S., {Sana}, H., \& {Sander}, A.~A.~C.
  2020, \aap, 634, A79

\bibitem[{Singh {et~al.}(2002)Singh, Naik, Paul, Agrawal, Rao, \&
  Singh}]{CygX-3_Singh2002}
Singh, N., Naik, S., Paul, B., {et~al.} 2002, Astronomy \& Astrophysics, 392,
  161

\bibitem[{{Spera} \& {Mapelli}(2017)}]{spera2017_pisnSNe}
{Spera}, M. \& {Mapelli}, M. 2017, \mnras, 470, 4739

\bibitem[{{Spera} {et~al.}(2015){Spera}, {Mapelli}, \&
  {Bressan}}]{Spera2015_remnantspectrum}
{Spera}, M., {Mapelli}, M., \& {Bressan}, A. 2015, \mnras, 451, 4086

\bibitem[{{Spera} {et~al.}(2019){Spera}, {Mapelli}, {Giacobbo}, {Trani},
  {Bressan}, \& {Costa}}]{spera2019_mergingBBH}
{Spera}, M., {Mapelli}, M., {Giacobbo}, N., {et~al.} 2019, \mnras, 485, 889

\bibitem[{{Sukhbold} {et~al.}(2018){Sukhbold}, {Woosley}, \&
  {Heger}}]{Sukhbold2018_compactnessnonmonotonic}
{Sukhbold}, T., {Woosley}, S.~E., \& {Heger}, A. 2018, \apj, 860, 93

\bibitem[{Van Der~Hucht(2001)}]{VanDerHucht2001_WRcatalogueMW}
Van Der~Hucht, K.~A. 2001, New Astronomy Reviews, 45, 135

\bibitem[{{van Son} {et~al.}(2022){van Son}, {de Mink}, {Renzo}, {Justham},
  {Zapartas}, {Breivik}, {Callister}, {Farr}, \&
  {Conroy}}]{VanSon2022_stableMTexplainsGWmass}
{van Son}, L.~A.~C., {de Mink}, S.~E., {Renzo}, M., {et~al.} 2022, \apj, 940,
  184

\bibitem[{Veledina {et~al.}(2024)Veledina, Muleri, Poutanen, Podgorn{\`y},
  Dov{\v{c}}iak, Capitanio, Churazov, De~Rosa, Di~Marco, Forsblom,
  {et~al.}}]{Veledina24_CygX3reviewULX}
Veledina, A., Muleri, F., Poutanen, J., {et~al.} 2024, Nature Astronomy, 1

\bibitem[{{Verbunt} {et~al.}(2017){Verbunt}, {Igoshev}, \&
  {Cator}}]{Verbunt2017_bimodalkicks}
{Verbunt}, F., {Igoshev}, A., \& {Cator}, E. 2017, \aap, 608, A57

\bibitem[{Vigna-G{\'o}mez {et~al.}(2020)Vigna-G{\'o}mez, MacLeod, Neijssel,
  Broekgaarden, Justham, Howitt, de~Mink, Vinciguerra, \&
  Mandel}]{VignaGomez2020_CEforBNS}
Vigna-G{\'o}mez, A., MacLeod, M., Neijssel, C.~J., {et~al.} 2020, Publications
  of the Astronomical Society of Australia, 37, e038

\bibitem[{Vink {et~al.}(2001)Vink, de~Koter, \& Lamers}]{Vink2001}
Vink, J.~S., de~Koter, A., \& Lamers, H. 2001, Astronomy \& Astrophysics, 369,
  574

\bibitem[{{Vink} {et~al.}(2000){Vink}, {de Koter}, \&
  {Lamers}}]{VinkdeKoterLamers2000_windsOB}
{Vink}, J.~S., {de Koter}, A., \& {Lamers}, H.~J.~G.~L.~M. 2000, \aap, 362, 295

\bibitem[{{Vink} {et~al.}(2001){Vink}, {de Koter}, \&
  {Lamers}}]{VinkdeKoterLamers2001_windsOB}
{Vink}, J.~S., {de Koter}, A., \& {Lamers}, H.~J.~G.~L.~M. 2001, \aap, 369, 574

\bibitem[{{Vink} {et~al.}(2011){Vink}, {Muijres}, {Anthonisse}, {de Koter},
  {Gr{\"a}fener}, \& {Langer}}]{Vink2011}
{Vink}, J.~S., {Muijres}, L.~E., {Anthonisse}, B., {et~al.} 2011, \aap, 531,
  A132

\bibitem[{{Wadekar} {et~al.}(2024){Wadekar}, {Venumadhav}, {Mehta}, {Roulet},
  {Olsen}, {Mushkin}, {Zackay}, \& {Zaldarriaga}}]{Wadekar2023_GWtemplates}
{Wadekar}, D., {Venumadhav}, T., {Mehta}, A.~K., {et~al.} 2024, \prd, 110,
  084035

\bibitem[{{Webbink}(1984)}]{Webbink1984_CE}
{Webbink}, R.~F. 1984, \apj, 277, 355

\bibitem[{{Willcox} {et~al.}(2023){Willcox}, {MacLeod}, {Mandel}, \&
  {Hirai}}]{Willcox2023_MTnonconservativeCOMPAS}
{Willcox}, R., {MacLeod}, M., {Mandel}, I., \& {Hirai}, R. 2023, \apj, 958, 138

\bibitem[{{Zdziarski} {et~al.}(2013){Zdziarski}, {Mikolajewska}, \&
  {Belczynski}}]{Cyg-X3_Zd2013}
{Zdziarski}, A.~A., {Mikolajewska}, J., \& {Belczynski}, K. 2013, \mnras, 429,
  L104

\end{thebibliography}

\appendix
\section{Models and assumptions}
\subsection{Fallback in the Mfb natal kick model}\label{app:fallback}
The Mfb natal kick model assigns the kick magnitudes drawing a random number $\rm f_{\rm \sigma}$ from a Maxwellian distribution ($\sigma=265 \kms$ 
rms) and modulating it with the fraction $f_{\rm fb, CCSN}$ of mass falling back onto the proto-NS:

\begin{equation}
    v_{\rm kick} = f_{\sigma=265} ~(1-f_{\rm fb, CCSN}).
\end{equation}

Here, we choose the fallback fraction $f_{\rm fb, CCSN}$ consistently with the CCSN models (Section \ref{subsec:CCSNmodels}). 

In the rapid and delayed case, we set the fallback fraction equal to the fallback parameter  $f_{\rm fb, CCSN} = f_{\rm fb}$ of \citep{Fryer2012}. In the compactness-based case, we distinguish successful and failed supernovae. For BHs formed via direct collapse, no material is ejected ($f_{\rm fb, CCSN} = 1$) and no kick is imparted. For NSs, we calculate a fallback fraction in analogy with \citet{Fryer2012}: we assume that the remnant mass ($\mrem$) is determined by the fraction $f_{\rm fb, CCSN}$ of the pre-SN mass ($M_{\rm pre-SN}$) that a proto-CO of $\mproto = 1.1 \msun$ is able to accrete, where 

\begin{equation}
    f_{\rm fb, CCSN}=\frac{\max(\mrem-\mproto,0)}{\max(\rm M_{\rm pre-SN}-\mproto,0)}.
\end{equation}

\subsection{Initial conditions for binary population}\label{app:initialcond}
The initial primary masses $M_1$ were drawn from the \citet{Kroupa2001} initial mass function (IMF) and the initial mass ratios $q$ from \citet{Sana2012}:

\begin{equation}
	\xi(M_1) \propto M_1^{-2.3} \qquad M_1 \in [5,150]~\msun,
\end{equation}
\begin{equation}
\xi(q) \propto q^{-0.1} \qquad q \in [0.1,1],
\end{equation}
where $q=M_2/M_1$ and $M_1 \geq M_2$. Combining $\xi(q)$ and $\xi(M_1)$, we obtained the distribution for the initial secondary masses $\xi(M_2) = \xi(M_1) \,{}\xi(q)$. We choose the lower cuts at $M_\textup{1,min} = 5~\msun$ and $q_\textup{min}=0.1$ because we focus on BH and NS progenitors. 

We generated a total mass of $M_{\rm IC} \approx 1.08 \times 10^7 \rm~ \msun$. Accounting for incomplete mass sampling of the primary from the Kroupa IMF requires a correction factor of $f_{\rm IMF} = 0.289$, indicating a parent population of $M_{\rm tot} \approx 3.7 \times 10^8 \rm~ \msun$.

We generated the initial orbital periods $P$ from the $\mathcal{P} = \log(P/\text{days})$ distribution by \citet{Sana2012}:

\begin{equation}
	\xi(\mathcal{P}) \propto \mathcal{P}^{-0.55} \qquad \mathcal{P} \in [0.30,5.5],
\end{equation}
setting $P=2~$days as lower limit (we assumed that binaries with shorter period already circularized following \citealt{MoeDiStefano2017}). \citet{MoeDiStefano2017} showed that there is a complex correlation between the mass ratio $q$, period $P$ and eccentricity $e$ of early-type binaries. For simplicity, we assumed independent distributions for $q$ and $P$ and we adopted the \citet{MoeDiStefano2017} prescription for $e$:

\begin{equation}
	\xi (e (P)) \propto 1-(P/\text{days})^{-2/3} \qquad P \geq 2~\text{d}.
\end{equation}

\section{Additional plots}\label{app:plotsextra}
\subsection{Formation channels and CE efficiency}\label{appsub:ace1}
\begin{figure*}
	\centering
 \includegraphics[width=.92\textwidth]{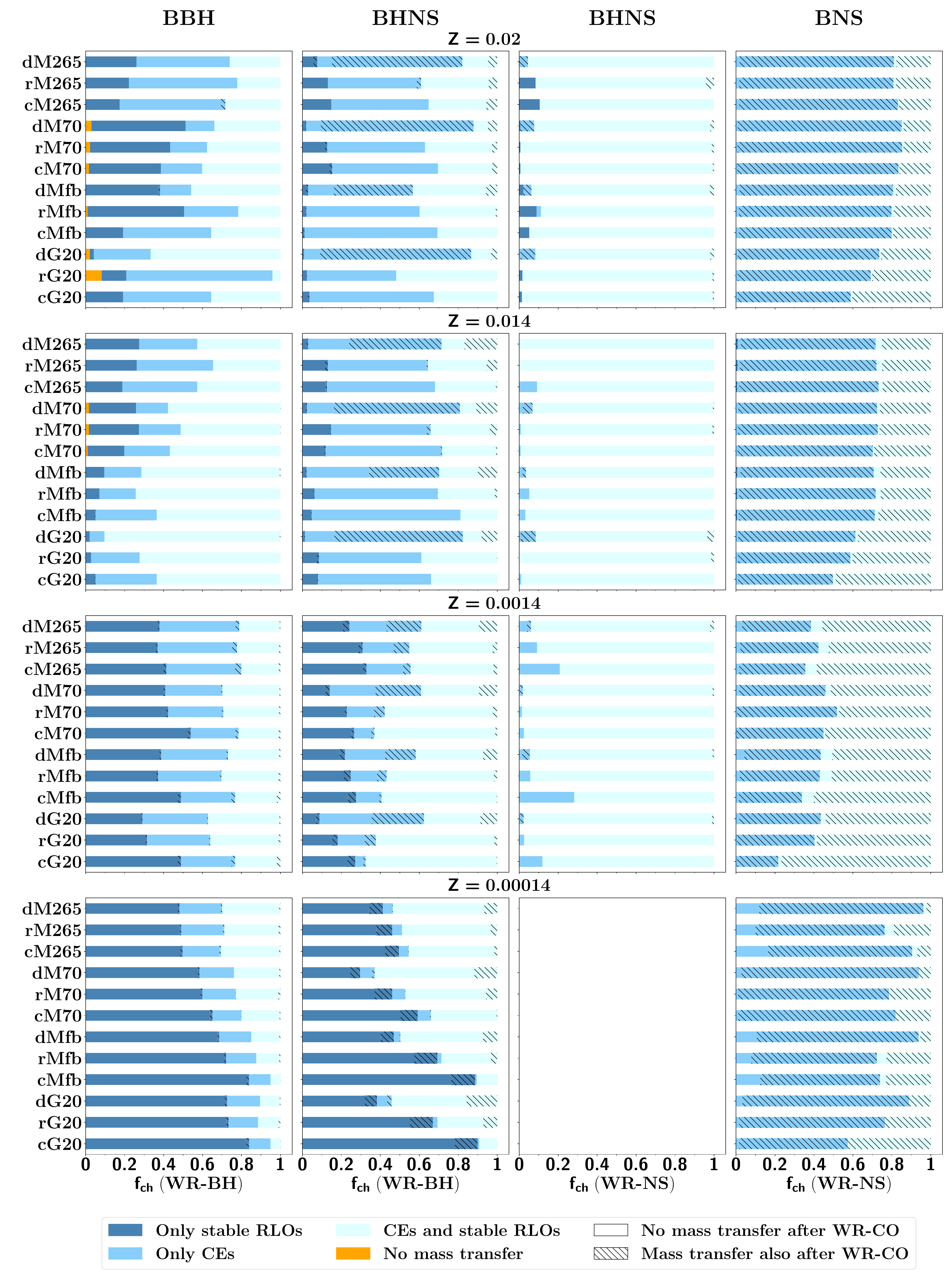}
 \caption{Fraction of BCOs with WR--CO progenitors evolved through different formation channels for  different metallicities (rows) and combinations of CCSN and natal kick models explored in this work (Table \ref{tab:parameterspace}). Here we show the $\alpha_{\rm CE}=1$ case. We distinguish four main evolutionary paths on the basis of mass transfer events occurred before the WR--CO formation: stable mass transfer only, CEs only, at least one stable and one unstable mass transfer, no mass transfer. Hatched bars indicate WR--CO systems that become BCOs experiencing at least one additional mass transfer event (stable or unstable RLO) after the WR--CO phase.}\label{fig:channels-a1}
\end{figure*}

In Figure \ref{fig:channels-a1} we show the formation channels of BCOs for \ace=1. The most notable difference with respect to the $\ace=3$ case (Figure \ref{fig:channels-a3}, Section \ref{subsec:formationchannels}) occurs for WR--NS formation and final fate. WR--NSs require one or more CE events to form, and at least an additional one to produce a BCO. At $Z=0.00014$, the role of CE evolution in BNS creation is enhanced at $\ace=1$, with $\approx 60-90 \%$ of BNSs exchanging mass only via CE events. At these metallicity, NSs may form only after efficient envelope stripping, as it is the one induced by CE evolution. Thus, evolution through stable RLOs is disfavoured However, a low CE efficiency determines inefficient CE ejection and can induce premature coalesce of the binary members. For instance, at $Z=0.00014$ and $\ace=1$ no WR--NS was able to survive the CE evolution and produce BHNSs, unlike the $\ace=3$ case.

\subsection{Orbital properties of Cyg X-3}\label{appsub:CygX3}
In Section \ref{subsec:CygX3}, we showed that Cyg X-3 is a possible BCO progenitor, with formation pathways representative of WR--COs evolved at $Z=0.02,0.014$. 

\begin{figure*}
        \centering
        \includegraphics[width=\textwidth]{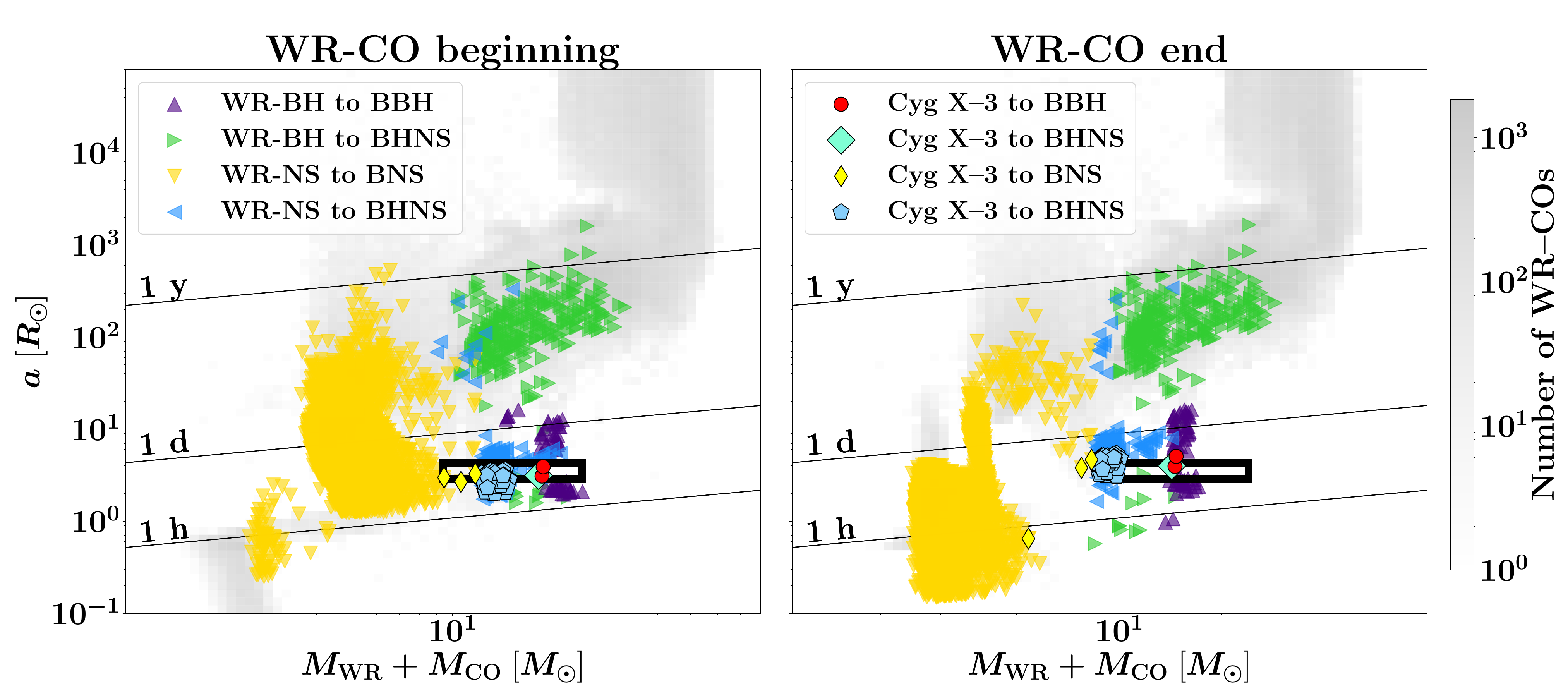}
	\caption{Orbital properties of WR--COs and Cyg X-3 at the beginning (left) and end (right) of the WR--CO phase. We highlight the properties of WR--COs (triangular markers) and Cyg X-3-like binaries (non-triangular markers) that are BCO progenitors, comparing them to the properties of the whole sample of WR--COs (gray areas in the background). We distinguish binaries ending up as BBHs (purple upward triangles; red circles), as BNSs (gold downward triangles; yellow thin diamonds), or as BHNSs, either forming the BH  (green rightward triangles; thick turquoise diamonds) or the NS first (blue leftward triangles; turquoise pentagons). The black rectangular box indicates the region of orbital parameters that we use to select Cyg X-3 candidates (Section \ref{subsec:CygX3}). Black diagonal lines indicate orbital periods of 1 hour, 1 day or 1 year. Here we show the set evolved with $\ace=3$, $Z=0.02$, G20 natal kick model and compactness CCSN model. In Figure \ref{fig:CygX3orbital} we show the same set, but with the delayed CCSN model.}\label{fig:CygX3orbital_compactness}
\end{figure*}

Figure \ref{fig:CygX3orbital_compactness} shows the orbital properties of WR--COs and Cyg X-3 in a set evolved with $\ace=3$, $Z=0.02$, the G20 natal kick model and the compactness-based CCSN model.  This binary population was evolved under the same assumptions of Figure \ref{fig:CygX3orbital}, except for the CCSN model. Despite the same \ace and $Z$ assumptions, the elevated probability of receiving a \emph{lucky kick} through the G20 model couples with the different CCSN model (compactness-based in Figure \ref{fig:CygX3orbital_compactness}, delayed in Figure \ref{fig:CygX3orbital}) and determines a different distribution of WR--COs in the space of orbital parameters. 

\begin{figure*}
        \centering
        \includegraphics[width=\textwidth]{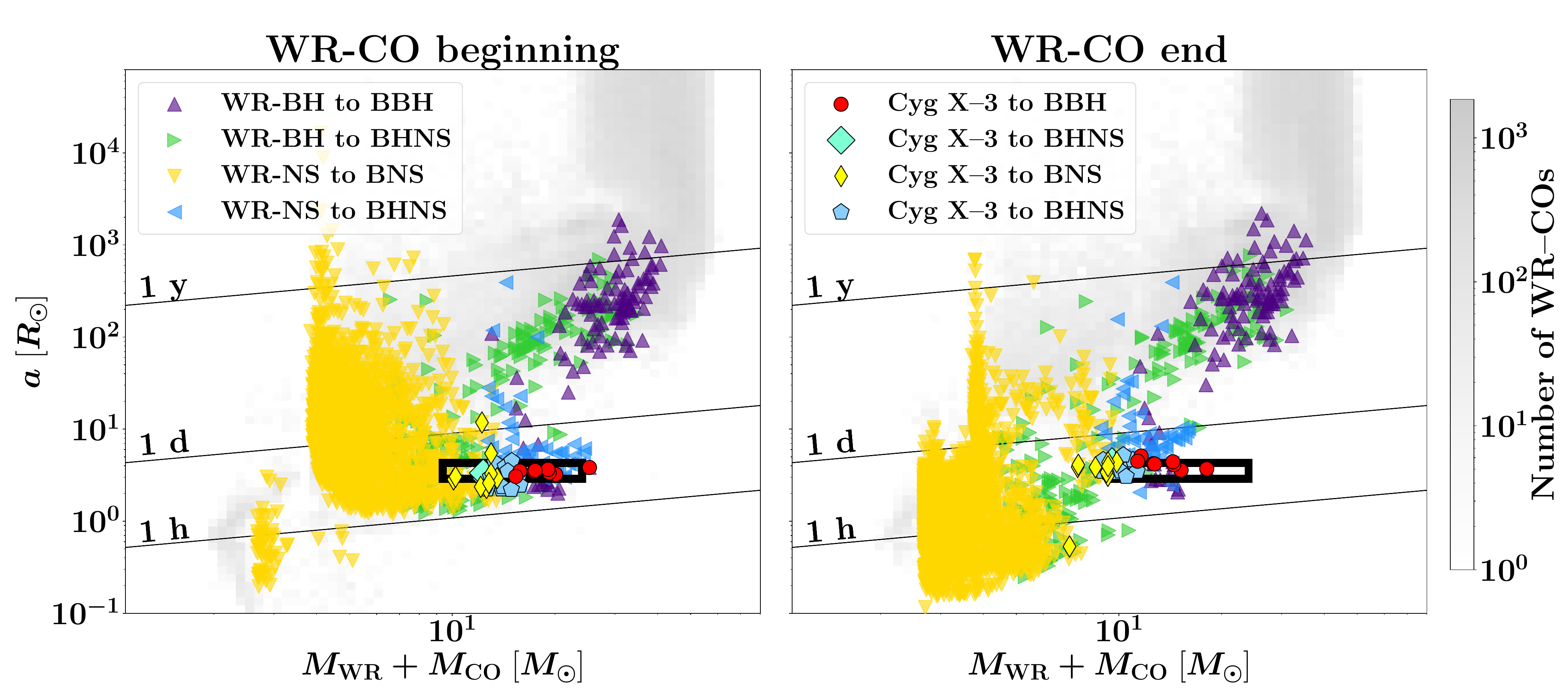}
	\caption{Same as Figure \ref{fig:CygX3orbital_compactness}, but here we show the set evolved with $\ace=3$, $Z=0.02$, Mfb natal kick model, and delayed CCSN model. In Figure \ref{fig:CygX3orbital} we show the same set, but with the G20 natal kick model.}\label{fig:CygX3orbital_Mfb}
\end{figure*}

For further comparison, we show in Figure \ref{fig:CygX3orbital_Mfb} how the occupation of space of orbital parameters changes if we assumed the kick model to be the Mfb one. In this plane, changing the initial models modifies the number of WR--COs with Cyg X-3-like properties. Nevertheless, binaries similar to Cyg X-3 remain favoured as BCO progenitors, suggesting that that further observations of WR--COs with orbital properties similar to Cyg X-3 may constrain the properties of BCO progenitors.

\end{document}